\documentclass[12pt]{article}
\usepackage{amssymb,amsmath,epsfig}
\usepackage{graphicx}
\usepackage[skip=0pt]{caption}
\usepackage{subcaption}
\usepackage{float}
\usepackage{epstopdf,leftidx,mathtools}
\usepackage{arydshln}

\newcommand{\be}{\begin{equation}}
\newcommand{\ee}{\end{equation}}
\newcommand{\bea}{\begin{eqnarray}}
\newcommand{\eea}{\end{eqnarray}}
\newcommand{\beaa}{\begin{eqnarray*}}
\newcommand{\eeaa}{\end{eqnarray*}}

\newcommand{\BB}{{{\rm I} \kern -2pt \rlap {\rm B} \kern +8pt}}

\DeclareCaptionFormat{myformat}{\fontsize{9}{10}\selectfont#1#2#3}
\advance\textheight by \topskip
\makeatletter

\@addtoreset{equation}{section}

\def\section{\@startsection {section}{1}{\z@}{-3.5ex plus -1ex minus
 -.2ex}{2.3ex plus .2ex}{\large\bf\centering}}
\def\subsection{\@startsection{subsection}{2}{\z@}{-3.25ex plus%
 -1ex minus -.2ex}{1.5ex plus .2ex}{\bf}}
\def\subsubsection{\@startsection{subsubsection}{3}{\z@}{-3.25ex plus%
 -1ex minus -.2ex}{1.5ex plus .2ex}{\sl}}
\makeatother
\begin{document}

\baselineskip 18pt
\parindent 12pt
\parskip 10pt

\begin{center}
{\large {\bf Integrable semi-discretization of complex and
multi-component coupled dispersionless systems and their
solutions}}\\\vspace{1
cm} { H. Wajahat A. Riaz \footnote{%
ahmed.phyy@gmail.com} and Mahmood ul Hassan \footnote{%
mhassan.physics@pu.edu.pk} }\vspace{0.15in}

{\small{\it Department of Physics, University of Punjab,
Quaid-e-Azam Campus, \\Lahore-54590, Pakistan.}}
\end{center}
\vspace{2cm}

\begin{abstract}
An integrable semi-discretization of complex and multi-component
coupled dispersionless systems via Lax pairs is presented. A Lax
pair is proposed for the complex sdCD system. We derive the Lax pair
for the multi-component sdCD system through generalizing the $2
\times 2$ Lax matrices to the case of $2^{N} \times 2^{N}$ Lax
matrices. A Darboux transformation (DT) is applied to the complex
and multi-component sdCD systems and is used to compute soliton
solutions of the systems. It is also shown that the soliton
solutions of the semi-discrete systems reduce to the continuous
systems by applying continuum limit.
\end{abstract}
\vspace{0.5 cm} \textit{PACS: 02.30.Ik, 05.45.Yv
\\Keywords: Discrete Integrable systems,
solitons, Darboux transformation}

\pagebreak
\section{Introduction}
Integrable discrete nonlinear equations defined on a lattice points
rather than a continuum space-time, play an important role in the
field of applied mathematics and nonlinear science. Generically,
they admit many solutions known as soliton solutions. Due to
solitonic behavior, integrable discrete systems become very
compelling from a physical point of view. Such systems are not only
useful in the study of numerical scheme but they also serve as
physical models in which space-coordinates are defined on lattice
points and time is taken as continuous. Examples are found in
various disciplines of science such as nonlinear lattices, plasma
physics, statistical mechanics and optical fibers etc.
\cite{a5}-\cite{riaz0}.

In the last couple of decades, dispersionless or quasiclassical
limits of integrable equations and hierarchies have received much
attention by researchers since they arise in the analysis of several
problems in applied mathematics and physics from the theory of
quantum fields and conformal maps on the complex plane
\cite{takasaki1}-\cite{konno1}. In \cite {konno}, some authors
proposed a set of coupled dispersionless (CD) integrable system
given by
\begin{eqnarray}
\partial_{x}\partial_{t}q + 2 \partial_{x}r r &=&0, \label{CD1} \\
\partial_{x}\partial_{t}r - 2 \partial_{x}q r &=&0. \label{CD2}
\end{eqnarray}
where $q \equiv q(x,\;t)$ and $r \equiv r(x,\;t)$ are real functions
of $x$ and $t$. The system (\ref{CD1})-(\ref{CD2}) is called
dispersionless because of not containing the dispersion term rather
than arising as a semi-classical limit. A generalized version of the
CD system was introduced by the same authors \cite{kakuhata} and is
given by
\begin{eqnarray}
\partial_{x}\partial_{t} q + \partial_{x}(rs) &=&0, \label{gCD1} \\
\partial_{x}\partial_{t}r - 2  \partial_{x}q r &=&0, \label{gCD2} \\
\partial_{x}\partial_{t}s - 2  \partial_{x}q s &=&0, \label{gCD3}
\end{eqnarray}
where $s\equiv s(x, t)$ is a real function of $x$ and $t$. The
inverse scheme of the set of equations (\ref{gCD1})-(\ref{gCD3}) is
given by
\begin{eqnarray}
\partial_{x}\Psi &=& U \Psi = -\dot\imath \lambda^{-1}\left(
\begin{array}{cc}
\partial_{x}q & \partial_{x}r \\
\partial_{x}s & -\partial_{x}q
\end{array}%
\right)\Psi , \label{glax1} \\
 \partial_{t}\Psi &=& V \Psi = \left[\left(
\begin{array}{cc}
0 & -r \\
s & 0
\end{array}%
\right) + \dot\imath \lambda \left(
\begin{array}{cc}
\frac{1}{2} & 0 \\
0 & -\frac{1}{2}
\end{array}%
\right)\right] \Psi. \label{glax2}
\end{eqnarray}
For $s = \bar{r}$ (where $\bar{r}$ represent complex conjugate of
$r$), we get the complex CD system \cite{konno1}. Physically CD
system describes the interaction of current-fed string in a certain
external magnetic field. CD system and complex CD system have been
shown to be gauge equivalent to the sine-Gordon equation and
Pohlmeyer-Lund-Regge model \cite{hirota}, \cite{kotlyarov}. Most
recently, the link of the motion of space curves to the real and
complex CD system and to the real and complex short-pulse equations
has been established via hodograph transformation. Further, the
connection is shown to be made clear between the real (complex) CD
system and real (complex) short-pulse equations and also with the
two-component Kadomtsew-Petviashvili (KP) hierarchy \cite{shen}. In
past years, the CD and complex CD systems, their generalization to a
system based on non-abelian Lie group and its non-commutative
extension have been studied and proved to be completely solvable by
inverse scattering method, Darboux transformation and with the help
of other methods of generating solutions \cite{konno}-\cite{konno1},
\cite{alagsan}-\cite{lou}.

Along with continuous coupled dispersionless system, its discrete
version also preserves the integrability of the system. In
\cite{vinet}-\cite{vinet1}, Vinet and Yu presented a discretization
of the CD and generalized CD system and obtained soliton solutions
via Hirota bilinear method. In \cite{riaz}-\cite{riaz2}, Darboux
transformations are used to discretize the CD system and their
generalizations. Meanwhile the behavior of soliton solutions have
also been investigated. In the present work, we present a
semi-discretization of the complex CD system via Lax pair. We shall
also study an integrable semi-discritization of multi-component
generalization of the complex CD system
\begin{eqnarray}
&& \partial_{x} \partial_{t} q +
\partial_{x}\left(\sum_{j=1}^{N}|r^{(j)}|^{2}\right) = 0,
\label{mCD1} \\
&&\partial_{x} \partial_{t}r^{(j)} - 2 q r^{(j)} = 0, \quad j =
1,\;2,\;3,\;...,\;N. \label{mCD2}
\end{eqnarray}
The multi-component complex CD system (\ref{mCD1})-(\ref{mCD2})
proposed in \cite{xu} and the soliton solutions were investigated by
means of Hirota bilinear method.

The paper is structured as follows. In section 2, we present an
integrable semi-discretization of the complex CD system via Lax
representation. We then extend the $2 \times 2$ Lax matrices for the
case of $2^{N} \times 2^{N}$ Lax matrices and obtain a
multi-component generalization of sdCD system. We also reduce both
complex sdCD and multi-component sdCD system to their respective
continuous counterparts by applying continuum limit. In section 3,
we define Darboux transformation (DT) for the multi-component
complex sdCD system and obtain multisoliton solutions. Further, we
present quasideterminant solutions of the multi-component sdCD
system. In section 4, we present soliton solutions for the complex
and 2-component complex sdCD system.

 \section{Lax pair representation}
We start with the Lax pair of the complex sdCD system which is a set
of differential-difference equations where space is taken as one
dimensional lattice and time is taken as continuous given by
\begin{eqnarray}
\Psi_{n+1} &=& \mathcal L_{n} \Psi_{n}= \left(
\begin{array}{cc}
1-\dot\imath \lambda^{-1}(q_{n+1}-q_{n})  & -\dot\imath \lambda^{-1}(r_{n+1}-r_{n}) \\
-\dot\imath \lambda^{-1}(\bar{r}_{n+1}-\bar{r}_{n}) & 1+\dot\imath
\lambda^{-1}(q_{n+1}-q_{n})
\end{array}%
\right), \label{scalarlax1} \\
\frac{d}{dt}\Psi_{n} &=& \mathcal M_{n}\Psi_{n} = \left(
\begin{array}{cc}
\frac{\dot\imath \lambda}{2}  & -r_{n} \\
\bar{r}_{n} & -\frac{\dot\imath \lambda}{2}
\end{array}%
\right), \label{scalarlax2}
\end{eqnarray}
where $n$ represents discrete index and $\bar{r}_{n}$ represents the
complex conjugate of $r_{n}$ and $\lambda$ is a spectral parameter.
The zero-curvature condition $\frac{d}{dt}\mathcal L_{n} + \mathcal
L_{n} \mathcal M_{n} - \mathcal M_{n+1} \mathcal L_{n} = 0$ yields
the complex sdCD system
\begin{eqnarray}
\frac{d}{dt}(q_{n+1}-q_{n}) + \left(\left|r_{n+1}\right|^{2} -
\left|r_{n}\right|^{2}\right) &=& 0, \label{scalarCD1} \\
\frac{d}{dt}(r_{n+1}-r_{n}) - (r_{n+1} + r_{n})(q_{n+1}-q_{n}) &=&
0. \label{scalarCD2}
\end{eqnarray}
The Lax pair for the multi-component generalization of the complex
sdCD system is obtained by extending the $2 \times 2$ Lax pair
matrices to the case of $2^{N} \times 2^{N}$ Lax matrices. The Lax
pair for the multi-component sdCD system is expressed as
\begin{eqnarray}
\Psi_{n+1}&=& \mathcal L_{n} \Psi_{n} = \left(\mathcal
J+\lambda^{-1}(\mathcal U_{n+1}-\mathcal U_{n})\right)\Psi_{n} ,
\label{lax1}
\\
\frac{d}{dt}\Psi_{n}&=&\mathcal M_{n} \Psi_{n} = \left(\mathcal
V_{n} + \lambda \mathcal V_{0}\right) \Psi_{n}, \label{lax2}
\end{eqnarray}
where $\mathcal J,\; \mathcal U_{n},\; \mathcal V_{n} $ and
$\mathcal V_{0}$ are the $2^{N} \times 2^{N}$ block matrices given
by
\begin{equation}
\mathcal J = \left( \begin{array}{cc}
I & O \\
O & I
\end{array}%
\right), \quad \mathcal U_{n} = -\dot{\imath}\left(
\begin{array}{cc}
\mathcal Q_{n} & \mathcal R_{n} \\
\mathcal S_{n} & -\mathcal Q_{n}
\end{array}%
\right), \quad \mathcal V_{n} = \left( \begin{array}{cc}
O & -\mathcal R_{n} \\
\mathcal S_{n} & O
\end{array}%
\right), \quad \mathcal V_{0} = \left( \begin{array}{cc}
\frac{\dot{\imath}}{2}I & O \\
O & -\frac{\dot{\imath}}{2}I
\end{array}%
\right), \label{matrices}
\end{equation}
where $\mathcal Q_{n}, \mathcal R_{n}$ and $\mathcal S_{n}=\mathcal
R_{n}^{\dag}$ (here $\dag$ in the superscript denotes Hermitian
conjugation) are the $2^{N-1} \times 2^{N-1}$ block matrices and
$O,\;I$ are the $2^{N-1} \times 2^{N-1}$ null and unit matrices,
respectively. The compatibility condition of the Lax pair
(\ref{lax1})-(\ref{lax2}) gives the matrix complex sdCD system
\begin{equation}
\frac{d}{dt}(\mathcal U_{n+1} - \mathcal U_{n}) + (\mathcal U_{n+1}
- \mathcal U_{n}) \mathcal V_{n} - \mathcal V_{n+1}(\mathcal U_{n+1}
- \mathcal U_{n}) = O. \label{matrixequation}
\end{equation}
By substituting the expression of $\mathcal U_{n}$ and $\mathcal
V_{n}$ from (\ref{matrices}) into (\ref{matrixequation}), we obtain
\begin{eqnarray}
\frac{d}{dt}(\mathcal Q_{n+1} - \mathcal Q_{n}) + \mathcal
R_{n+1}\mathcal S_{n+1} - \mathcal R_{n} \mathcal S_{n} &=& O,
\label{sdmCD1} \\
\frac{d}{dt}(\mathcal R_{n+1} - \mathcal R_{n}) - (\mathcal
Q_{n+1}-\mathcal Q_{n})\mathcal R_{n} - \mathcal R_{n+1} (\mathcal
Q_{n+1}-\mathcal Q_{n}) &=& O. \label{sdmCD2}
\end{eqnarray}
In the case, where we define the matrix variables $\mathcal
Q_{n},\;\mathcal R_{n}$ in terms of scalar dependent variables
$q_{n},\;r_{n}$, i.e. $\mathcal Q_{n}\equiv q_{n},\;\mathcal
R_{n}\equiv r_{n}$, one can reduce the matrix complex sdCD system
(\ref{sdmCD1})-(\ref{sdmCD2}) to the complex sdCD system
(\ref{scalarCD1})-(\ref{scalarCD2}), and when the matrices $\mathcal
Q_{n}$ and $\mathcal R_{n}$ take the form
\begin{eqnarray}
&&\mathcal Q_{n}^{(2)} = q_{n} I_{2 \times 2}, \quad {\mathcal
R_{n}^{(2)}} = \left(
\begin{array}{c:c}
r_{n}^{(1)} & r_{n}^{(2)} \\
\hdashline - \bar{r}_{n}^{(2)} & \bar{r}_{n}^{(1)}
\end{array}%
\right)_{2 \times 2}, \label{2by2} \\
&&\mathcal Q_{n}^{(3)}= q_{n}I_{4 \times 4} , \quad \mathcal
R_{n}^{(3)} = \left(
\begin{array}{c:c}
\begin{array}{cc}
  r_{n}^{(1)} & r_{n}^{(2)} \\
  -\bar{r}_{n}^{(2)} & \bar{r}_{n}^{(1)}
\end{array} & \begin{array}{cc}
  r_{n}^{(3)} & 0 \\
  0 & r_{n}^{(3)}
              \end{array} \\ \hdashline
              \begin{array}{cc}
  -\bar{r}_{n}^{(3)} & 0 \\
  0 & -\bar{r}_{n}^{(3)}
              \end{array} & \begin{array}{cc}
  \bar{r}_{n}^{(1)} & -r_{n}^{(2)} \\
  \bar{r}_{n}^{(2)} & r_{n}^{(1)}
  \end{array}
\end{array} \right)_{4 \times 4}, \label{4by4} \\
&&\mathcal Q_{n}^{(4)} = q_{n}I_{8 \times 8}, \label{8by8} \\
&& \mathcal R_{n}^{(4)} =\left(
\begin{array}{c:c}
\begin{array}{cccc}
  r_{n}^{(1)} & r_{n}^{(2)} & 0 & 0 \\
  -\bar{r}_{n}^{(2)} & \bar{r}_{n}^{(1)} & 0 & 0 \\
  0 & 0 & r_{n}^{(1)} & r_{n}^{(2)} \\
  0 & 0 & -\bar{r}_{n}^{(2)} & \bar{r}_{n}^{(1)}
\end{array} & \begin{array}{cccc}
  r_{n}^{(3)} & 0 & r_{n}^{(4)} & 0 \\
  0 & r_{n}^{(3)} & 0 & r_{n}^{(4)} \\
  -\bar{r}_{n}^{(4)} & 0 & \bar{r}_{n}^{(3)} & 0 \\
  0 & -\bar{r}_{n}^{(4)} & 0 & \bar{r}_{n}^{(3)}
              \end{array} \\ \hdashline
              \begin{array}{cccc}
  -\bar{r}_{n}^{(3)} & 0 & r_{n}^{(4)} & 0 \\
  0 & -\bar{r}_{n}^{(3)} & 0 & r_{n}^{(4)} \\
  -\bar{r}_{n}^{(4)} & 0 & -r_{n}^{(3)} & 0 \\
  0 & -\bar{r}_{n}^{(4)} & 0 & -r_{n}^{(3)}
              \end{array} & \begin{array}{cccc}
  \bar{r}_{n}^{(1)} & -r_{n}^{(2)} & 0 & 0 \\
  \bar{r}_{n}^{(2)} & r_{n}^{(1)} & 0 & 0 \\
  0 & 0 & \bar{r}_{n}^{(1)} & -r_{n}^{(2)} \\
  0 & 0 & \bar{r}_{n}^{(2)} & r_{n}^{(1)}
\end{array}
\end{array} \right)_{8 \times 8}, \notag
\end{eqnarray}
we get multi-component complex sdCD system. By substituting the
expressions of (\ref{2by2})-(\ref{8by8}) into the set of equations
(\ref{sdmCD1})-(\ref{sdmCD2}), we obtain respectively, the
2-component, 3-component and 4-component complex sdCD system. The
2-component complex sdCD is
\begin{eqnarray}
&&\frac{d}{dt}(q_{n+1}-q_{n}) +
\sum_{j=1}^{2}\left(|r_{n+1}^{(j)}|^{2}-|r_{n}^{(j)}|^{2}\right)=0,
\label{twocomponent1} \\
&&\frac{d}{dt}(r_{n+1}^{(j)}-r_{n}^{(j)}) -
(r_{n+1}^{(j)}+r_{n}^{(j)})(q_{n+1}-q_{n})=0, \quad j=1,\;2.
\label{twocomponent2}
\end{eqnarray}
The 3-component complex sdCD system is
\begin{eqnarray}
&&\frac{d}{dt}(q_{n+1}-q_{n}) +
\sum_{j=1}^{3}\left(|r_{n+1}^{(j)}|^{2}-|r_{n}^{(j)}|^{2}\right)=0,
 \\
&&\frac{d}{dt}(r_{n+1}^{(j)}-r_{n}^{(j)}) -
(r_{n+1}^{(j)}+r_{n}^{(j)})(q_{n+1}-q_{n})=0, \quad j=1,\;2,\;3.
\label{threecomponent}
\end{eqnarray}
Similarly, $N$-component complex sdCD system is given by
\begin{eqnarray}
&&\frac{d}{dt}(q_{n+1}-q_{n}) +
\sum_{j=1}^{N}\left(|r_{n+1}^{(j)}|^{2}-|r_{n}^{(j)}|^{2}\right)=0,
\label{ncomp1} \\
&&\frac{d}{dt}(r_{n+1}^{(j)}-r_{n}^{(j)}) -
(r_{n+1}^{(j)}+r_{n}^{(j)})(q_{n+1}-q_{n})=0, \quad j=1,\;...,\;N.
\label{ncomp2}
\end{eqnarray}
In general, the expressions of the $2^{N-1} \times 2^{N-1}$ matrices
$\mathcal Q_{n}^{(N)}$ and $\mathcal R_{n}^{(N)}$ are
\begin{eqnarray}
\mathcal Q_{n}^{(N)}=q_{n}I_{2^{N-1} \times 2^{N-1}}, \quad \mathcal
R_{n}^{(N)} = \left(
\begin{array}{c:c}
\mathbb{R}_{n}^{(1)} & \mathbb{R}_{n}^{(2)} \\
\hdashline \mathbb{R}_{n}^{(3)} & \mathbb{R}_{n}^{(4)}
\end{array}%
\right), \label{general}
\end{eqnarray}
where $\mathbb{R}_{n}^{(j)}$ ($j=1,\;2,\;3,\;4$) are all $2^{N-2}
\times 2^{N-2}$ square block matrices, and
$\mathbb{R}_{n}^{(4)}=(\mathbb{R}_{n}^{(1)})^{\dag},\;\mathbb{R}_{n}^{(3)}=-(\mathbb{R}_{n}^{(2)})^{\dag}$.
The matrices $\mathbb{R}_{n}^{(1)}$ and $\mathbb{R}_{n}^{(2)}$ are
given by
\begin{equation}\label{xymatrix}
\mathbb{R}_{n}^{(1)} = \left(
\begin{array}{ccccc}
\mathbb{X}_{n}^{(1)} & & & \text{\Huge O} & \\
& \mathbb{X}_{n}^{(1)} & & & \\
& & \ddots & & \\
&  &  & \mathbb{X}_{n}^{(1)}& \\
& \text{\Huge O} & & & \mathbb{X}_{n}^{(1)}
\end{array}%
\right), \quad \mathbb{R}_{n}^{(2)} = \left(
\begin{array}{c:c}
\mathbb{Y}_{n}^{(1)} & \mathbb{Y}_{n}^{(2)} \\
\hdashline \mathbb{Y}_{n}^{(3)} & \mathbb{Y}_{n}^{(4)}
\end{array}%
\right),
\end{equation}
where
$\mathbb{Y}_{n}^{(4)}=(\mathbb{Y}_{n}^{(1)})^{\dag},\;\mathbb{Y}_{n}^{(3)}=-(\mathbb{Y}_{n}^{(2)})^{\dag}$
are all square block matrices which take values in the following
form
\begin{eqnarray}
&&\mathbb{Y}_{n}^{(1)} = \left(
\begin{array}{cccc}
\mathbb{X}_{n}^{(2)} & & & \text{\Huge O}  \\
& \mathbb{X}_{n}^{(2)}  & & \\
& & \ddots &  \\
 \text{\Huge O}& & &  \mathbb{X}_{n}^{(2)}
\end{array}%
\right), \; \mathbb{Y}_{n}^{(2)} = \left(
\begin{array}{ccccc}
\mathbb{X}_{n}^{(3)} & \mathbb{O} & \hdots & \mathbb{X}_{n}^{(N-1)} & \mathbb{X}_{n}^{(N)} \\
\mathbb{O} & \mathbb{X}_{n}^{(3)}  & \hdots &-\mathbb{X}_{n}^{(N)} & \mathbb{X}_{n}^{(N-1)} \\
\vdots & \vdots & \ddots & \vdots & \vdots \\
-\mathbb{X}_{n}^{(N-1)}& \mathbb{X}_{n}^{(N)} & \hdots & \mathbb{X}_{n}^{(3)} & \mathbb{O} \\
-\mathbb{X}_{n}^{(N)} & -\mathbb{X}_{n}^{(N-1)} & \hdots &
\mathbb{O} & \mathbb{X}_{n}^{(3)}
\end{array}%
\right), \notag \\
&& \mathbb{O} = \left(
\begin{array}{cc}
0 & 0 \\
0 & 0
\end{array}%
\right), \quad \mathbb{X}_{n}^{(1)} = \left(
\begin{array}{cc}
r_{n}^{(1)} & r_{n}^{(2)} \\
-\bar{r}_{n}^{(2)} & \bar{r}_{n}^{(1)}
\end{array}%
\right),  \quad \mathbb{X}_{n}^{(j)} = \left(
\begin{array}{cc}
r_{n}^{(j+1)} & 0 \\
0 & r_{n}^{(j+1)}
\end{array}%
\right), \notag \\ && \qquad \qquad \qquad \qquad \qquad \qquad
\qquad \qquad \qquad \qquad \qquad  j=2,\;3,\;...,\;N-1.
\label{y2y3}
\end{eqnarray}
Substituting (\ref{general}) and (\ref{y2y3}) into the set of
equations (\ref{sdmCD1})-(\ref{sdmCD2}), we obtain $N$-component
complex sdCD system (\ref{ncomp1})-(\ref{ncomp2}). Equations
(\ref{scalarCD1})-(\ref{scalarCD2}) and
(\ref{ncomp1})-(\ref{ncomp2}) represents respectively, an integrable
semi-discrete (discrete in space-coordinate) analogue of the complex
and multi-component complex sdCD system. They can be reduced to the
continuous complex and multi-component complex CD systems by
applying continuum limit. For this, let us define $\lim_{\delta
\rightarrow 0}\frac{f_{n+1} - f_{n}}{\delta} = f_{x}$, where
$\delta$ is the lattice parameter in space-direction. Applying this
to equations (\ref{ncomp1})-(\ref{ncomp2}), one can obtain
multi-component complex CD system given by
\begin{eqnarray}
&& \partial_{x} \partial_{t} q +
\partial_{x}\left(\sum_{j=1}^{N}|r^{(j)}|^{2}\right) = 0,
\label{contmCD1} \\
&&\partial_{x} \partial_{t}r^{(j)} - 2 q r^{(j)} = 0, \quad j =
1,\;2,\;3,\;...,\;N. \label{contmCD2}
\end{eqnarray}
For $N=1$, the set of equations (\ref{contmCD1})-(\ref{contmCD2})
reduce to the usual complex CD system
\begin{eqnarray}
&& \partial_{x} \partial_{t} q + \partial_{x}|r|^{2} = 0,
\label{usualCD1} \\
&&\partial_{x} \partial_{t}r - 2 q r = 0. \label{usualCD2}
\end{eqnarray}
\section{Darboux transformation}
Darboux transformation (DT) is one of solution generating technique
in soliton theory that allow us to express the solutions for a given
integrable equation in simple explicit form
\cite{15}-\cite{riazmSP}. In this section, we construct the DT of
the multi-component complex sdCD system
(\ref{sdmCD1})-(\ref{sdmCD2}).

In what follows, we shall apply a DT to the solutions of the Lax
pair and the solutions of the multi-component complex sdCD system
and then express in terms of quasideterminents. Let us define a new
solution $\Psi_{n}^{[1]}$ to the Lax pair (\ref{lax1})-(\ref{lax2})
which is related to the old solution $\Psi_{n}$ by means of $2^{N}
\times 2^{N}$ matrix $ \mathcal D_{n} (t; \; \lambda)$ called the
Darboux matrix. The one-fold DT on the solution to the Lax pair is
given by
\begin{equation}\label{d1}
\Psi_{n}^{[1]}=\mathcal D_{n}(t; \; \lambda)\Psi_{n}=\left(\lambda
\mathcal I-\Xi_{n}\right)\Psi_{n}.
\end{equation}
In the present case, $\mathcal I$ is a $2^{N} \times 2^{N}$ unit
matrix and $\Xi_{n}$ is an invertible $2^{N} \times 2^{N}$ matrix,
to be determined. The new solution $\Psi_{n}^{[1]}$ satisfies the
same Lax pair i.e.
\begin{eqnarray}
\Psi_{n+1}^{[1]} &=& \mathcal L_{n}^{[1]} \Psi_{n}^{[1]} =
\left(\mathcal J^{[1]}+\lambda^{-1}(\mathcal U_{n+1}^{[1]}-\mathcal
U_{n}^{[1]})\right)\Psi_{n}^{[1]} , \label{dlax1}
\\
\frac{d}{dt}\Psi_{n}^{[1]} &=& \mathcal M_{n}^{[1]} \Psi_{n}^{[1]} =
\left(\mathcal V_{n}^{[1]} + \lambda \mathcal V_{0}^{[1]}\right)
\Psi_{n}^{[1]}. \label{dlax2}
\end{eqnarray}
By substituting the expression of $\Psi_{n}^{[1]}$ from (\ref{d1})
into Lax pair equations (\ref{dlax1})-(\ref{dlax2}), we obtain DT on
the matrices $\mathcal U_{n},\;\mathcal V_{n}, \; \mathcal{J}$ and
$\mathcal V_{0}$
\begin{eqnarray}
\mathcal U_{n}^{[1]} &=& \mathcal U_{n} - \Xi_{n}, \label{S1} \\
\mathcal V_{n}^{[1]} &=& \mathcal V_{n} + \left[\mathcal
V_{0},\;\Xi_{n}\right], \label{W1} \\
\mathcal J^{[1]} &=& \mathcal J, \quad \mathcal V_{0}^{[1]} =
\mathcal V_{0},
\end{eqnarray}
with the conditions on the matrix $\Xi_{n}$ arising due to the
Darboux covariance as follows
\begin{eqnarray}
\left(\Xi_{n+1} - \Xi_{n}\right)\Xi_{n} &=& \left(\mathcal U_{n+1} -
\mathcal U_{n}\right) \Xi_{n} - \Xi_{n+1}\left(\mathcal U_{n+1} -
\mathcal U_{n}\right), \label{1st condition} \\
\frac{d}{dt}\Xi_{n} &=& \left[\mathcal V_{n}, \; \Xi_{n}\right] +
\left[\mathcal V_{0},\;\Xi_{n}\right] \Xi_{n}. \label{2nd condition}
\end{eqnarray}
Now we construct the matrix $\Xi_{n}$ in terms of solutions to the
Lax pair (\ref{lax1})-(\ref{lax2}). For this we proceed as follows:

Define $2^{N}$ distinct constant parameters
$\lambda_{1},\;\lambda_{2},\;...,\;\lambda_{2^{N}}$ such that for
each $\lambda_{j}$ we have a peculiar column vector solution
$\left|f^{(j)}\right\rangle_{n}=\Psi(\lambda_{j})\left|e_{j}\right\rangle$
(where $\left|e_{j}\right\rangle$ is a constant column vector) to
the Lax pair (\ref{lax1})-(\ref{lax2}). For $\lambda = \lambda_{j}$
($j=1,\;2,\;...,\;2^{N}$), we write
\begin{eqnarray}
\left|f^{(j)}\right\rangle_{n+1} &=& \left|f^{(j)}\right\rangle_{n}
+ \lambda^{-1}_{j} \left(\mathcal
U_{n+1} - \mathcal U_{n}\right) \left|f^{(j)}\right\rangle_{n}, \label{column1} \\
\frac{d}{dt}\left|f^{(j)}\right\rangle_{n} &=& \mathcal V_{n}
\left|f^{(j)}\right\rangle_{n} + \lambda_{j} \mathcal V_{0}
\left|f^{(j)}\right\rangle_{n}. \label{column2}
\end{eqnarray}
Define $2^{N} \times 2^{N}$ constant eigenvalue matrix with entries
$\lambda_{j}$ i.e. $\Lambda = \text{diag}(\lambda_{1},\; ...,\;
\lambda_{2^{N}})$, and construct an invertible $2^{N} \times 2^{N}$
matrix $\mathcal F_{n}$ as
\begin{equation}
\mathcal F_{n} =
\left(\Psi_{n}(\lambda_{1})\left|e_{1}\right\rangle,\;...,\;
\Psi_{n}(\lambda_{2^{N}})\left|e_{2^{N}}\right\rangle\right) =
\left(\left|f^{(1)}\right\rangle_{n},\;...,\;
|f^{(2^{N})}\rangle_{n}\right),
\end{equation}
so that the Lax pair (\ref{lax1})-(\ref{lax2}) with the particular
matrix $\mathcal F_{n}$ as solution, can be written as
\begin{eqnarray}
\mathcal F_{n+1}&=& \mathcal F_{n} + \left(\mathcal U_{n+1} -
\mathcal U_{n}\right)\mathcal F_{n}\Lambda^{-1}, \label{matrixlax1}
\\
\frac{d}{dt}\mathcal F_{n} &=& \mathcal V_{n} \mathcal F_{n} +
\mathcal V_{0} \mathcal F_{n} \Lambda. \label{matrixlax2}
\end{eqnarray}
Further we check that the choice of the matrix $\Xi_{n} = \mathcal
F_{n} \Lambda \mathcal F_{n}^{-1}$ satisfy the conditions (\ref{1st
condition})-(\ref{2nd condition}) as imposed by Darboux covariance.
This can be checked by a direct computation as follows
\begin{eqnarray}
\left(\Xi_{n+1} - \Xi_{n}\right)\Xi_{n} &=& \mathcal F_{n+1} \Lambda
\mathcal F_{n+1}^{-1} \mathcal F_{n} \Lambda \mathcal F_{n}^{-1} -
\mathcal F_{n} \Lambda \mathcal F_{n}^{-1}\mathcal F_{n} \Lambda
\mathcal F_{n}^{-1} \notag \\
&+& \mathcal F_{n+1} \Lambda \mathcal F_{n}^{-1}\mathcal F_{n}
\Lambda \mathcal F_{n}^{-1} - \mathcal F_{n+1} \Lambda \mathcal
F_{n+1}^{-1}\mathcal F_{n+1} \Lambda \mathcal F_{n}^{-1}, \notag \\
&=& \left(\mathcal F_{n+1} \Lambda \mathcal F_{n}^{-1} - \mathcal
F_{n} \Lambda \mathcal F_{n}^{-1}\right)\mathcal F_{n} \Lambda
\mathcal F_{n}^{-1} - \mathcal F_{n+1} \Lambda \mathcal F_{n+1}^{-1}
\left(\mathcal F_{n+1} \Lambda \mathcal F_{n}^{-1} - \mathcal F_{n}
\Lambda \mathcal F_{n}^{-1}\right), \notag \\
&=& \left(\mathcal U_{n+1} - \mathcal U_{n}\right)\Xi_{n} -
\Xi_{n+1} \left(\mathcal U_{n+1} - \mathcal U_{n}\right), \\
\frac{d}{dt}\Xi_{n} &=& \left(\frac{d\mathcal F_{n}}{dt}\right)
\Lambda \mathcal F_{n}^{-1} - \mathcal F_{n} \Lambda \mathcal
F_{n}^{-1}\left(\frac{d\mathcal F_{n}}{dt}\right) \mathcal
F_{n}^{-1}, \notag \\
&=& \left(\mathcal V_{n} \mathcal F_{n} + \mathcal V_{0} \mathcal
F_{n} \Lambda\right)\Lambda \mathcal F_{n}^{-1} - \mathcal F_{n}
\Lambda \mathcal F_{n}^{-1}\left(\mathcal V_{n} \mathcal F_{n} +
\mathcal V_{0} \mathcal F_{n} \Lambda\right) \mathcal
F_{n}^{-1}, \notag \\
&=& \left[\mathcal V_{n}, \; \Xi_{n}\right] + \left[\mathcal
V_{0},\;\Xi_{n}\right] \Xi_{n}.
\end{eqnarray}
So the conditions (\ref{1st condition}) and (\ref{2nd condition})
are satisfied. Hence we have established that one-fold DT to the
solutions of the Lax pair and the solutions of the multi-component
complex sdCD system can be expressed as
\begin{eqnarray}
\Psi_{n}^{[1]} &=& \mathcal D_{n}(t; \; \lambda) \Psi_{n} = (\lambda
\mathcal I - \mathcal F_{n} \Lambda \mathcal F_{n}^{-1}) \Psi_{n}, \label{1foldpsi}\\
\mathcal U_{n}^{[1]} &=& \mathcal U_{n} - \mathcal F_{n} \Lambda
\mathcal F_{n}^{-1}. \label{1folds}
\end{eqnarray}
The one-fold DT $\Psi_{n}^{[1]}$ and $\mathcal U_{n}^{[1]}$ can be
written in terms of quasideterminants \footnote{We use the notion of
quasideterminants which have various useful properties that play
important roles in constructing exact solutions of integrable
equations. For details (see e.g. \cite{gelfand1})} and then by
$K$-times iteration process, one can obtain the $K$-fold DT by using
the properties of quaisdeterminants. For the Darboux matrix
$\mathcal D_{n}(t; \; \lambda) = \lambda \mathcal I - \mathcal F_{n}
\Lambda \mathcal F_{n}^{-1}$, equations (\ref{1foldpsi}) and
(\ref{1folds}) can be expressed as
\begin{eqnarray}
\Psi_{n}^{[1]}&=&{D_{n}(t; \; \lambda)}\Psi_{n}=\lambda\Psi_{n}-\mathcal F_{n} \Lambda \mathcal F_{n}^{-1} \Psi_{n}, \notag \\
&=&\lambda \Psi_{n} + \left \vert \begin{array}{cc}
\mathcal F_{n} & \Psi_{n} \\
\mathcal F_{n} \Lambda & \fbox{$O $}%
\end{array}%
\right\vert =\left\vert
\begin{array}{cc}
\mathcal F_{n} & \Psi_{n} \\
\mathcal F_{n} \Lambda & \fbox{$\lambda \Psi_{n}$}
\end{array}%
\right\vert,\label{d12} \\ \mathcal U_{n}^{[1]} &=& \mathcal U_{n} -
\mathcal F_{n} \Lambda \mathcal F_{n}^{-1}, \notag
\\
&=& \mathcal U_{n} + \left\vert
\begin{array}{cc}
\mathcal F_{n} & I \\
\mathcal F_{n} \Lambda & \fbox{$O $}%
\end{array}%
\right\vert.\label{d13}
\end{eqnarray}
For $\Lambda=\Lambda_{k}$ $\left(k=1,\;2,\;...,\;K\right)$, one can
write one-fold DT $\Psi_{n}^{[1]}$ to $K$-fold DT on the solution by
an iteration process, so that the $K$-fold DT $\Psi_{n}^{[K]}$ can
be expressed as
\begin{eqnarray}
\Psi_{n}^{[K]} &=&\left\vert
\begin{array}{ccccc}
\mathcal F_{n,\;1} & \mathcal F_{n,\;2} & \cdots & \mathcal F_{n,\;K} & \Psi_{n} \\
\mathcal F_{n,\;1}\Lambda_{1} & \mathcal F_{n,\;2}\Lambda_{2} & \cdots & \mathcal F_{n,\;K}\Lambda_{K} & \lambda \Psi_{n} \\
\vdots & \vdots & \ddots & \vdots & \vdots\\
\mathcal F_{n, \;1}\Lambda^{K}_{1} & \mathcal F_{n,
\;2}\Lambda^{K}_{2} & \cdots & \mathcal F_{n, \;K}\Lambda^{K}_{K} &
\fbox{$\lambda^{K} \Psi_{n}$}
\end{array}%
\right\vert.\label{d17}
\end{eqnarray}
Similarly the $K$-fold DT on the matrix $\mathcal U_{n}$ is
\begin{equation}\label{d18}
\mathcal U_{n}^{[K]} = \mathcal U_{n} +  \left\vert
\begin{array}{ccccc}
\mathcal F_{n, \;1} & \mathcal F_{n,\;2} & \cdots & \mathcal F_{n, \;K} & O \\
\mathcal F_{n, \;1}\Lambda_{1}   & \mathcal F_{n,\;2} \Lambda_{2} & \cdots & \mathcal F_{n,\;K} \Lambda_{K} & O \\
\vdots & \vdots & \ddots & \vdots & \vdots \\
\mathcal F_{n, \;1}\Lambda_{1}^{K-1} & \mathcal F_{n,\;2}
\Lambda_{2}^{K-1} &
\cdots & \mathcal F_{n,\;K} \Lambda_{K}^{K-1} & I \\
\mathcal F_{n,\;1}\Lambda_{1}^{K} & \mathcal F_{n, \;2}
\Lambda_{2}^{K} &
\cdots & \mathcal F_{n, \;K} \Lambda_{K}^{K} & \fbox{$ O $}%
\end{array}%
\right\vert.
\end{equation}
It appears to be more convenient to express the equation (\ref{d18})
in the following way
\begin{eqnarray}
\mathcal U_{n}^{[K]} &=& \mathcal U_{n}+\Theta_{n}^{[K]},
\label{SS1}
\end{eqnarray}
where $2^N \times 2^N$ matrix $\Theta_{n}^{[K]}$ is a
quasiedeterminant given by
\begin{equation}\label{S2}
\Theta_{n}^{[K]}=\left\vert
\begin{array}{cc}
\mathcal F_{n} & \mathcal E^{(K)} \\
\mathcal {\widehat F}_{n} & \fbox{$ O $}
\end{array}%
\right \vert,
\end{equation}
where $\mathcal E^{(K)}$ are $2^{N}K \times 2^{N}$ and $\mathcal
{\widehat F}_{n},\; \mathcal F_{n}$ are the $ 2^{N} \times 2^{N}K,\;
2^{N}K \times 2^{N}K$ matrices respectively, i.e.
\begin{eqnarray}
\mathcal E^{(K)}&=&\left(\begin{array}{cccc}
              O & O & \cdots & I
            \end{array}\right)^{T}, \notag \\
\mathcal {\widehat F}_{n}&=&\left(\begin{array}{cccc} \mathcal
F_{n,\;1}\Lambda^{K}_{1} & \mathcal F_{n,\;2} \Lambda^{K}_{2} &
\cdots & \mathcal F_{n,\;K} \Lambda^{K}_{K}
\end{array}\right),\notag \\
\mathcal F_{n}&=&\left(
  \begin{array}{cccc}
    \mathcal F_{n,\;1} & \mathcal F_{n,\;2} & \cdots & \mathcal F_{n,\;K} \\
    \mathcal F_{n,\;1} \Lambda_{1} & \mathcal F_{n,\;2} \Lambda_{2} & \cdots & \mathcal F_{n,\;K}\Lambda_{K} \\
    \vdots & \vdots & \ddots & \vdots \\
    \mathcal F_{n,\;1} \Lambda^{K-1}_{1} & \mathcal F_{n,\;2} \Lambda^{K-1}_{2} & \cdots & \mathcal F_{n,\;K} \Lambda^{K-1}_{K} \\
  \end{array}
\right).
\end{eqnarray}
The matrix elements of the matrix $\Theta_{n}^{[K]}$ can be computed
as
\begin{eqnarray}
\left({\Theta}_{n}^{[K]}\right)_{ij} &=& \left(\left\vert
\begin{array}{cc}
\mathcal F_{n} & \mathcal{E}^{(K)} \\
\mathcal {\widehat F}_{n}  & \fbox{$ O $}%
\end{array}%
\right\vert\right)_{ij} = \left\vert
\begin{array}{cc}
\mathcal F_{n} & \mathcal {E}^{(K)}_{j} \\
(\mathcal {\widehat F}_{n})_{i}  & \fbox{$ 0 $}%
\end{array}%
\right\vert, \quad i \neq j, \notag \\
\left({\Theta}_{n}^{[K]}\right)_{ij} &=& \left(\left\vert
\begin{array}{cc}
\mathcal F_{n} & \mathcal {E}^{(K)} \\
\mathcal {\widehat F}_{n}  & \fbox{$ O $}%
\end{array}%
\right\vert\right)_{ii} = \left\vert
\begin{array}{cc}
\mathcal F_{n} & \mathcal {E}^{(K)}_{i} \\
(\mathcal {\widehat F}_{n})_{i}  & \fbox{$ 0 $}%
\end{array}%
\right\vert, \quad i = j. \label{T1}
\end{eqnarray}
where $(\mathcal {\widehat F}_{n})_i$ indicates the $i$-th row of
$\mathcal F_{n}$ and $(\mathcal {E}^{(K)})_{j}$ represents $j$-th
column of $\mathcal{E}^{(K)}$ respectively. From the set of
equations (\ref{SS1})-(\ref{T1}) one can compute explicit
expressions of the DT on the scalar solutions of the complex and
multi-component sdCD systems.

\section{Soliton solutions}
In this section, we consider the complex and 2-component complex
sdCD system respectively and obtain one, two and three-soliton
solutions. In order to generate soliton solutions, let us take
matrix valued seed solution as
\begin{equation}\label{matrixseed}
\mathcal Q_{n+1} - \mathcal Q_{n} = \left(
\begin{array}{c:c}
\varrho I & O \\
\hdashline O & \varrho I
\end{array}%
\right), \quad \mathcal R_{n} = \left(
\begin{array}{c:c}
O & O \\
\hdashline O & O
\end{array}%
\right),
\end{equation}
where $\varrho$ is a non-zero real constant, so the matrix valued
solution $\Psi_{n}$ to the Lax pair (\ref{lax1})-(\ref{lax2}) can be
written as
\begin{equation}\label{matrixseed}
\Psi_{n} = \left(
\begin{array}{c:c}
\mathbb A_{n}(\lambda) & O \\
\hdashline O & \mathbb B_{n}(\lambda)
\end{array}%
\right)
\end{equation}
where $\mathbb A_{n}(\lambda) = \left(1-\dot{\imath}\varrho
\lambda^{-1}\right)^{n}e^{\frac{\dot\imath \lambda}{2}t} I_{2^{N-1}
\times 2^{N-1}}$ and $\mathbb B_{n}(\lambda) =
\left(1+\dot{\imath}\varrho
\lambda^{-1}\right)^{n}e^{-\frac{\dot\imath \lambda}{2}t} I_{2^{N-1}
\times 2^{N-1}}$. By using properties of quasideterminants in the
Darboux matrix, we obtain one-, two- and three-fold DT on the
solutions of the complex and 2-component complex sdCD systems in
terms of ratios of simple determinants.
\subsection{Complex sdCD system}
\begin{figure}[H]
        \centering
        \begin{subfigure}{0.32\textwidth}
                \includegraphics[width=\textwidth]{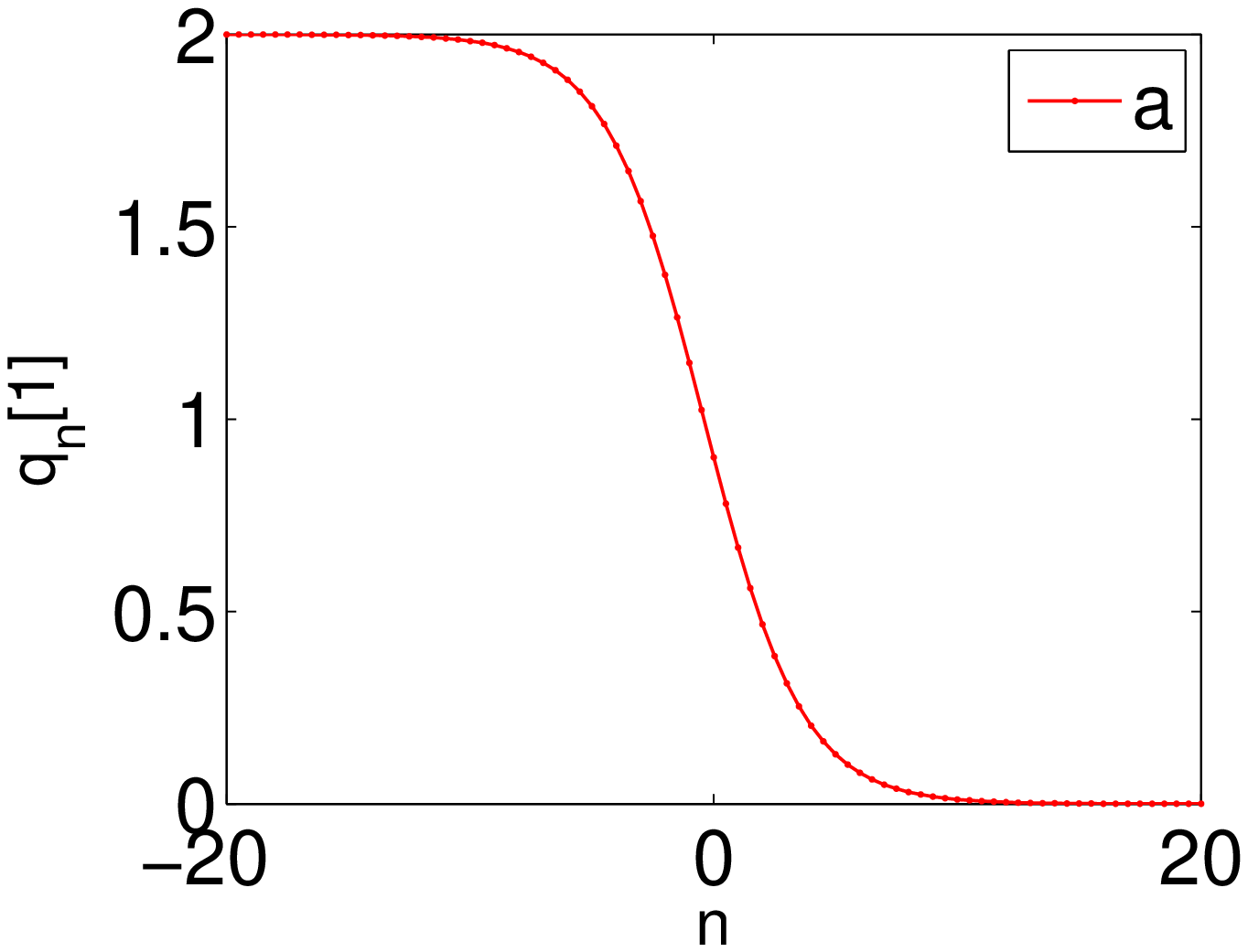}
                \label{fig:fig1}
        \end{subfigure}
        \begin{subfigure}{0.32\textwidth}
                \includegraphics[width=\textwidth]{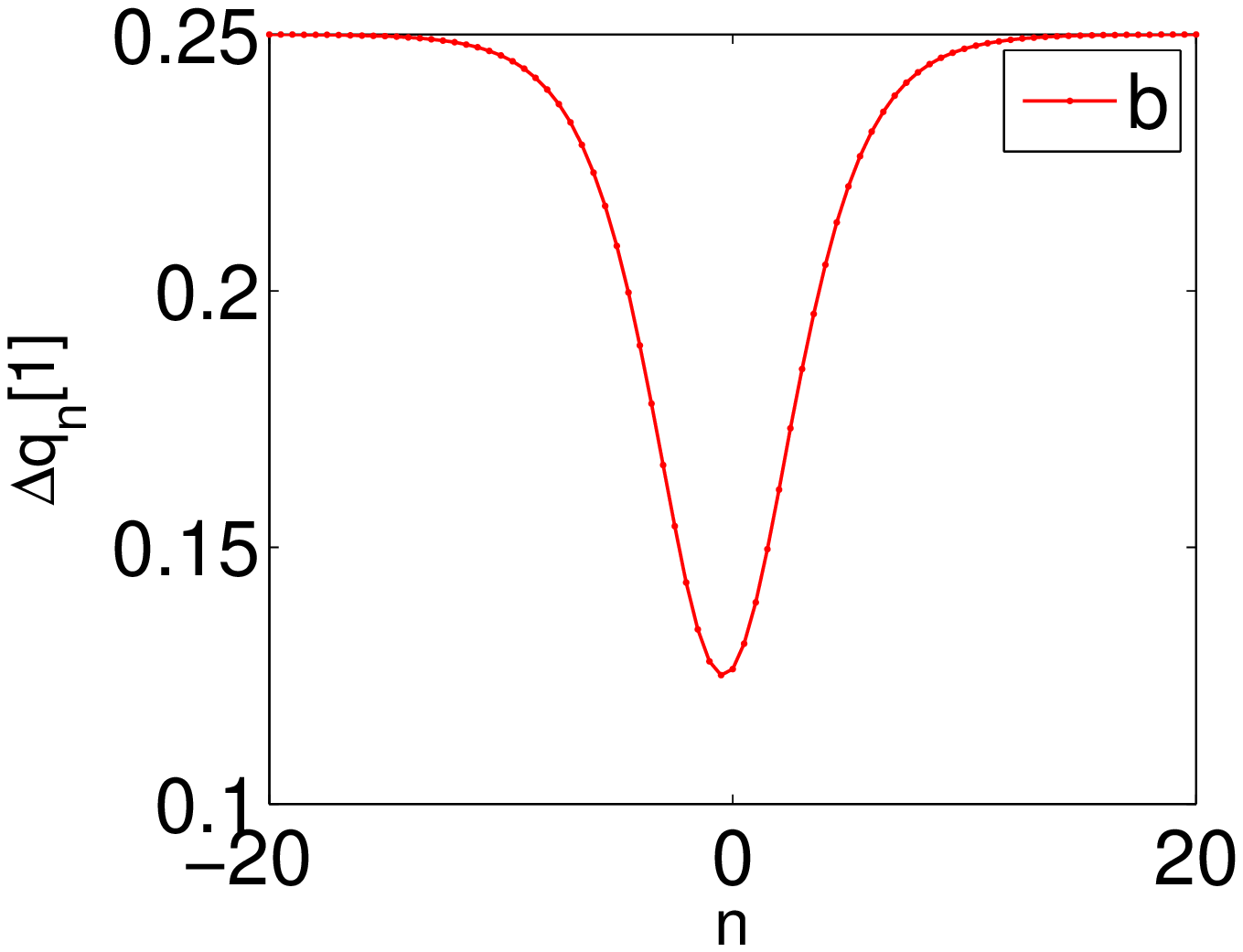}
                \label{fig:fig2}
        \end{subfigure}
        \begin{subfigure}{0.32\textwidth}
                \includegraphics[width=\textwidth]{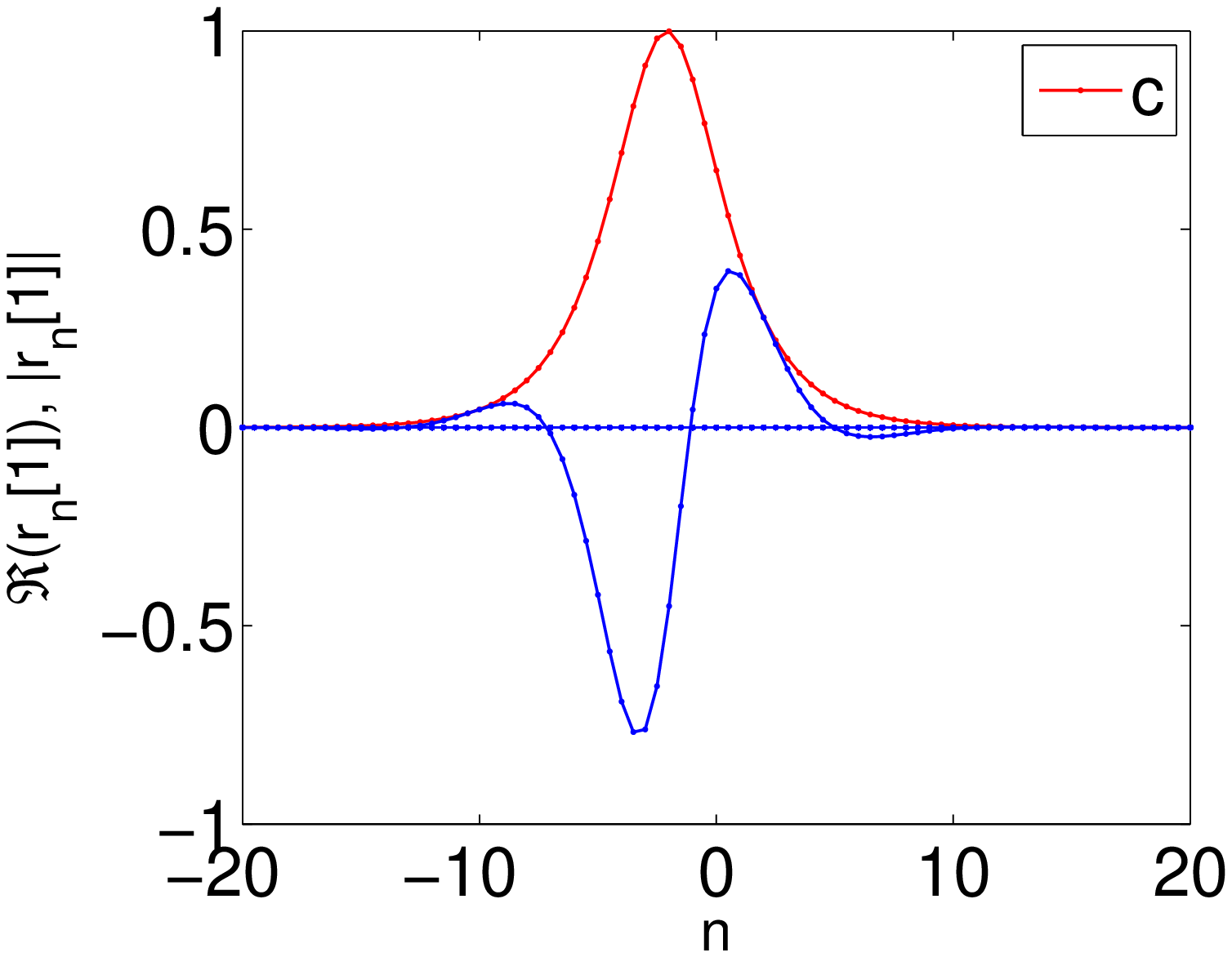}
                \label{fig:fig1}
        \end{subfigure}
        \label{fig:sd1}
        \caption{One-(kink and dark) soliton solutions in figures (a and b) respectively, with the choice of parameters
        $t=0.1,\;\varrho=0.25,\;\lambda_{1}=1+\dot\imath$ and in figure (c); red line: $|r_{n}[1]|$, blue line $\Re{(r_{n}[1])}$  with
        $t=1,\;\varrho=0.5,\;\lambda_{1}=1+\dot\imath$.}
        \label{fig:figure1}
\end{figure}

\begin{figure}[H]
        \centering
        \begin{subfigure}{0.32\textwidth}
                \includegraphics[width=\textwidth]{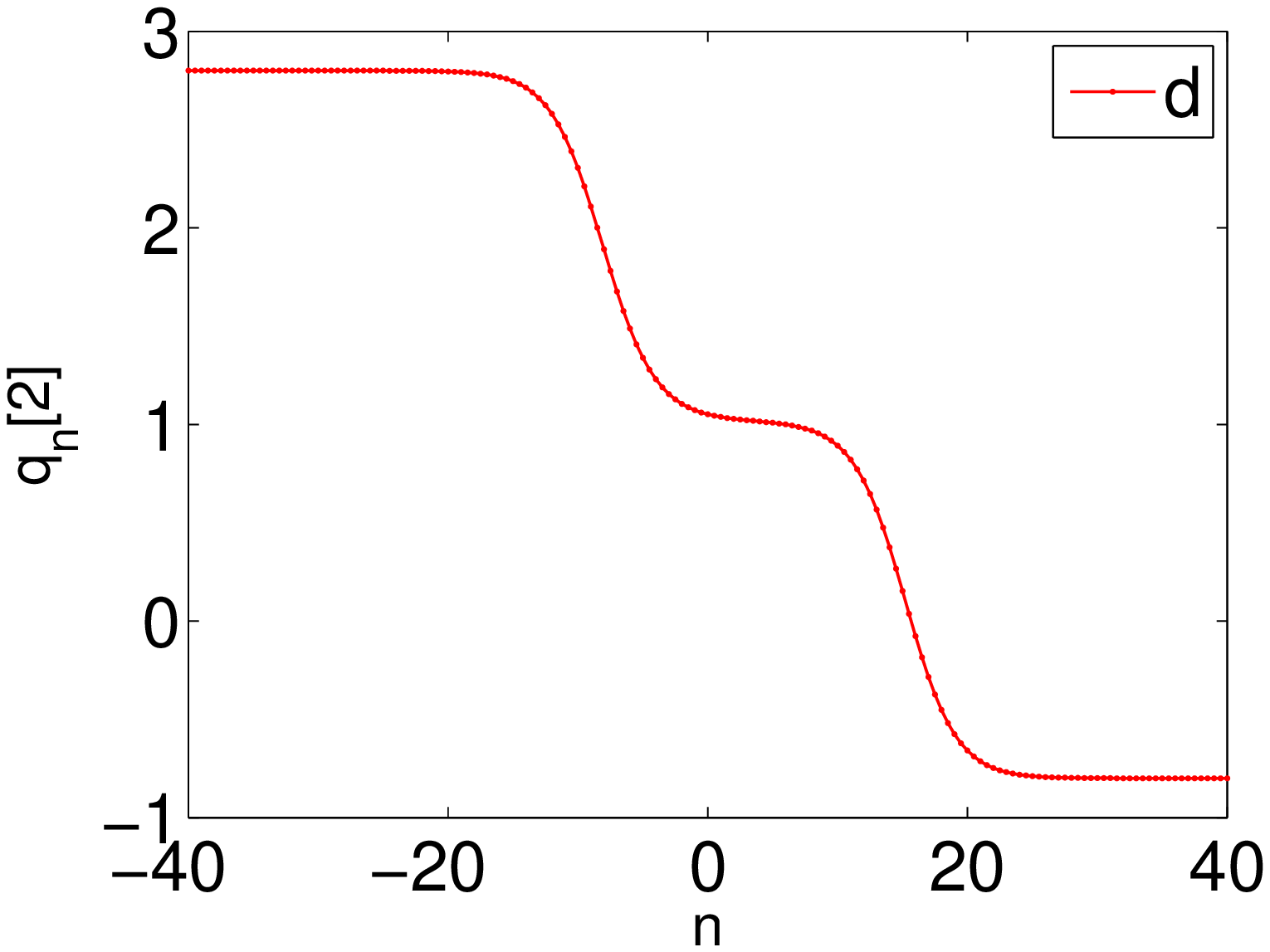}
                \label{fig:fig1}
        \end{subfigure}
        \begin{subfigure}{0.32\textwidth}
                \includegraphics[width=\textwidth]{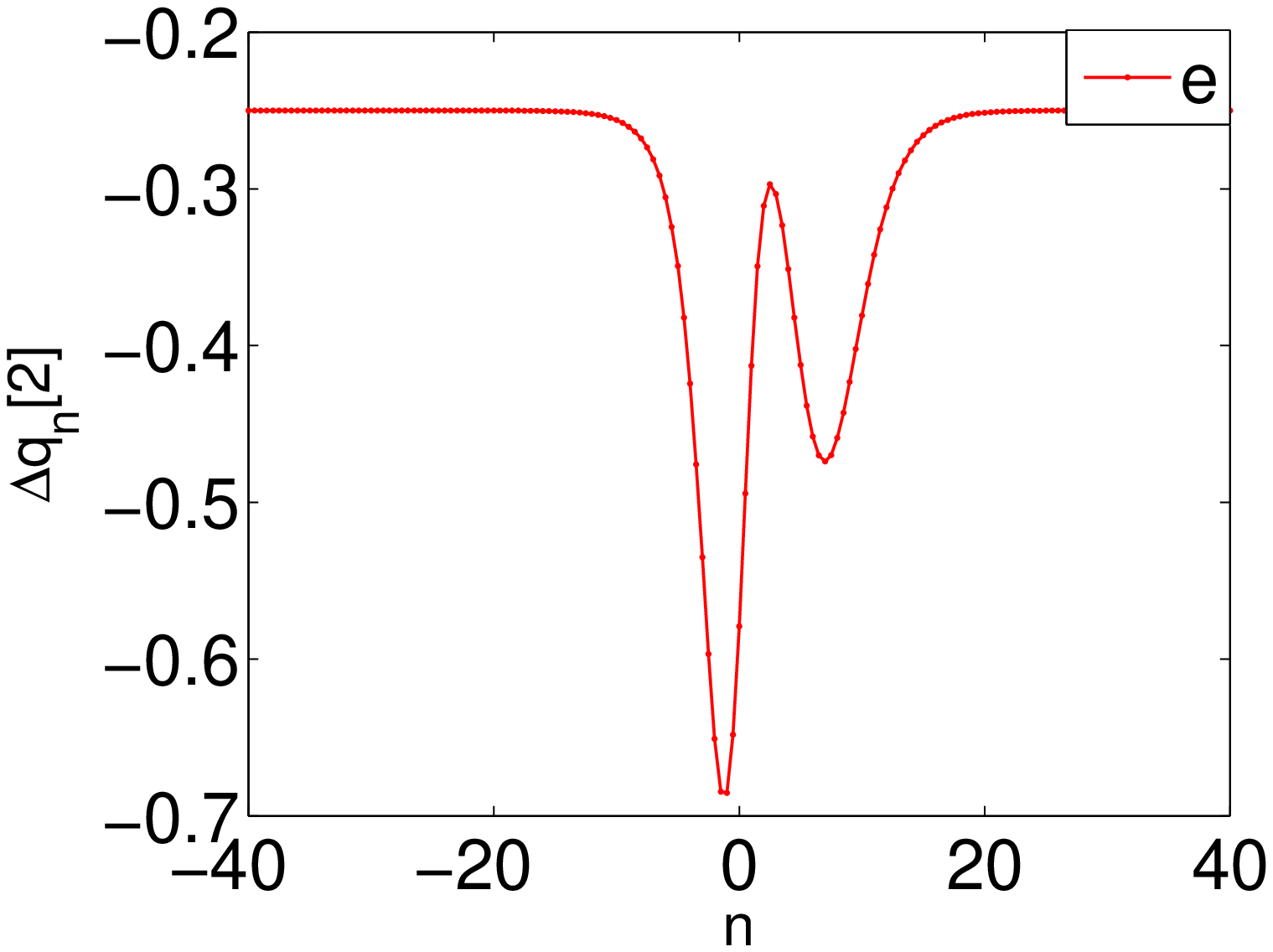}
                \label{fig:fig2}
        \end{subfigure}
        \begin{subfigure}{0.32\textwidth}
                \includegraphics[width=\textwidth]{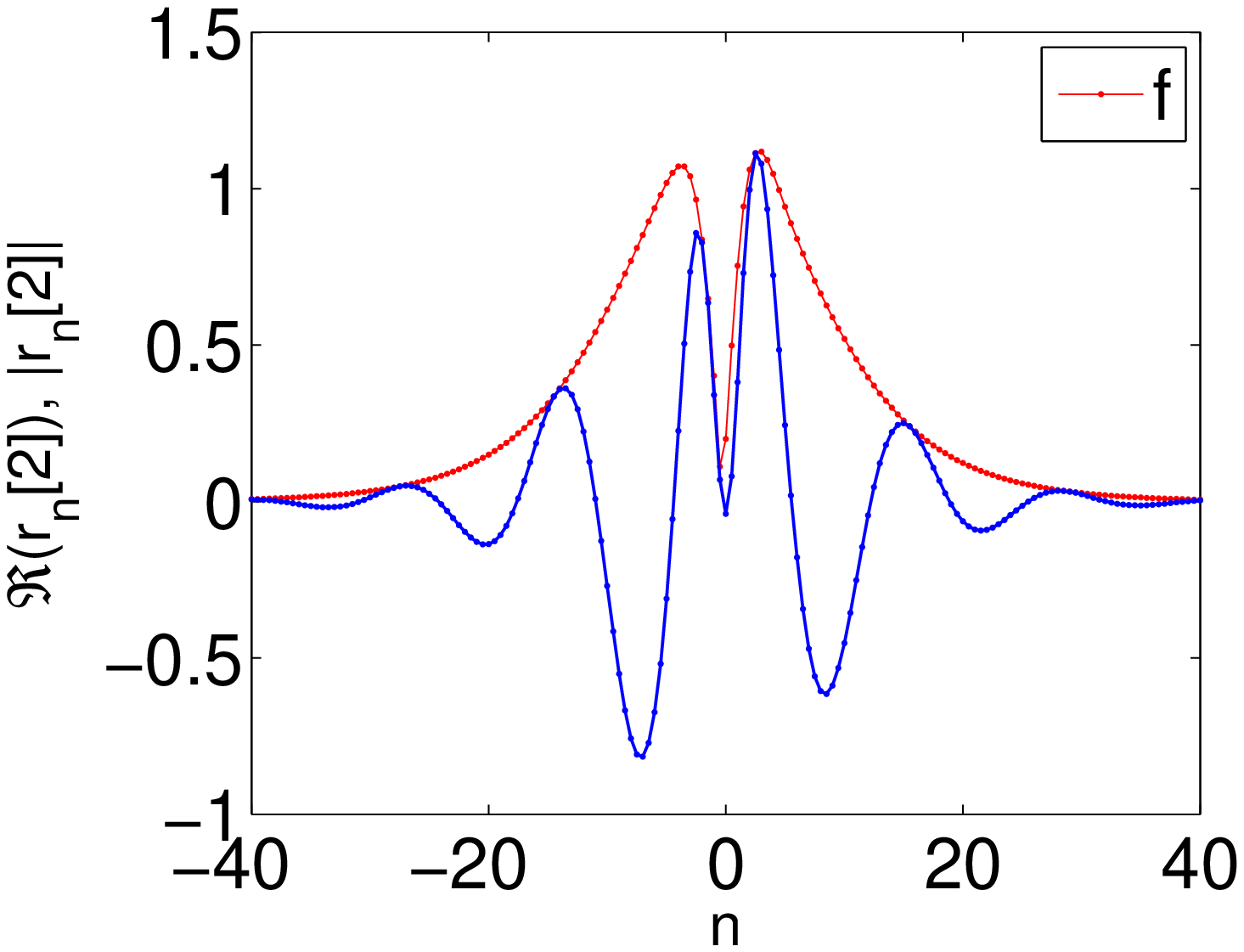}
                \label{fig:fig1}
        \end{subfigure}
        \caption{Two-(kink and dark) soliton solutions in figures (d and e) respectively, with the choice of parameters
        $t=1,\;\varrho=-0.25,\;\lambda_{1}=0.1+0.8\dot\imath,\;\lambda_{2}=1-\dot\imath$ and in figure (f); red line: $|r_{n}[2]|$, blue line $\Re{(r_{n}[2])}$  with
        $t=0.1,\;\varrho=0.8,\;\lambda_{1}=1+\dot\imath,\; \lambda_{2}=3-\dot\imath$.}
        \label{fig:figure2}
\end{figure}

\begin{figure}[H]
        \centering
        \begin{subfigure}{0.32\textwidth}
                \includegraphics[width=\textwidth]{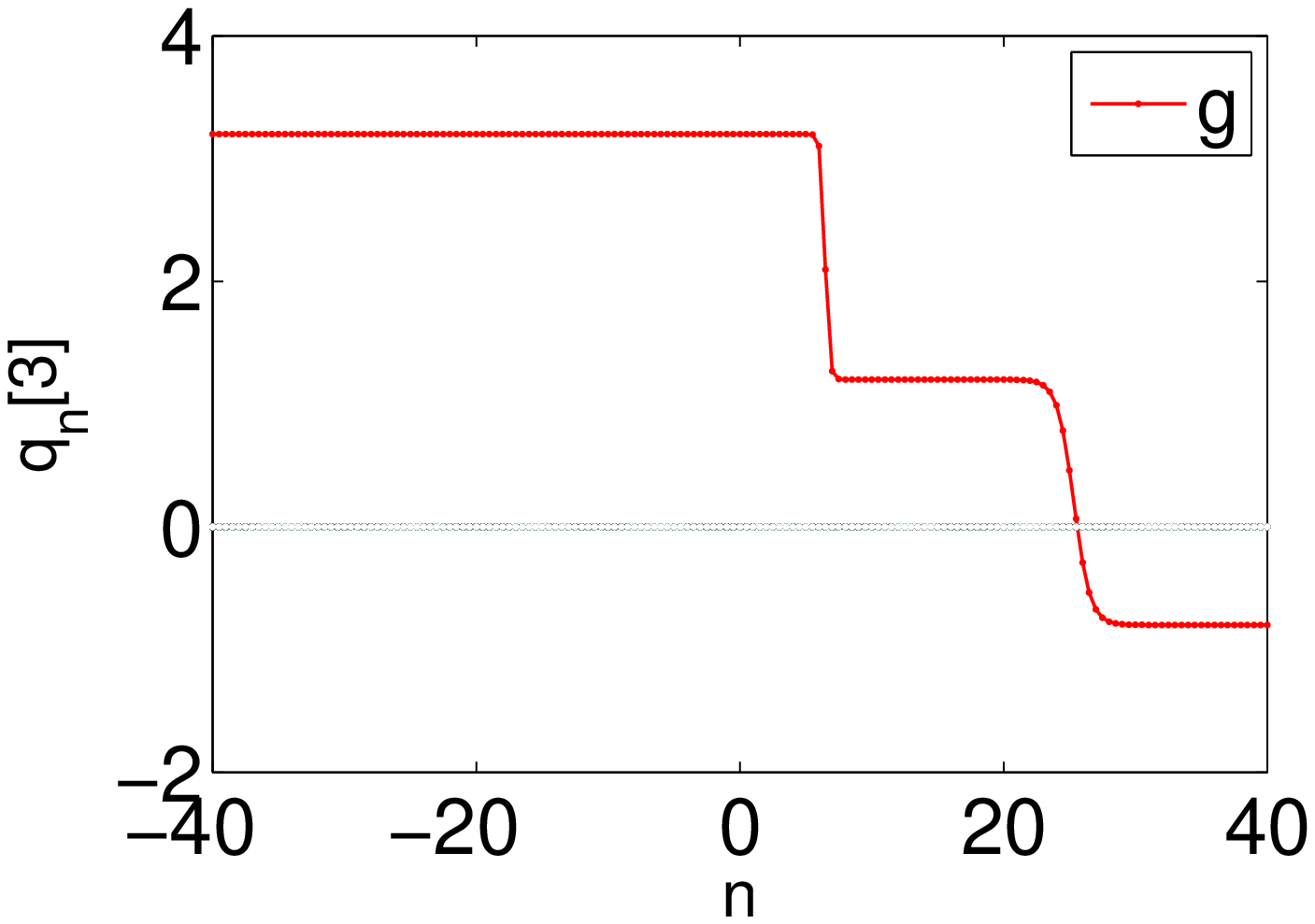}
                \label{fig:fig1}
        \end{subfigure}
        \begin{subfigure}{0.32\textwidth}
                \includegraphics[width=\textwidth]{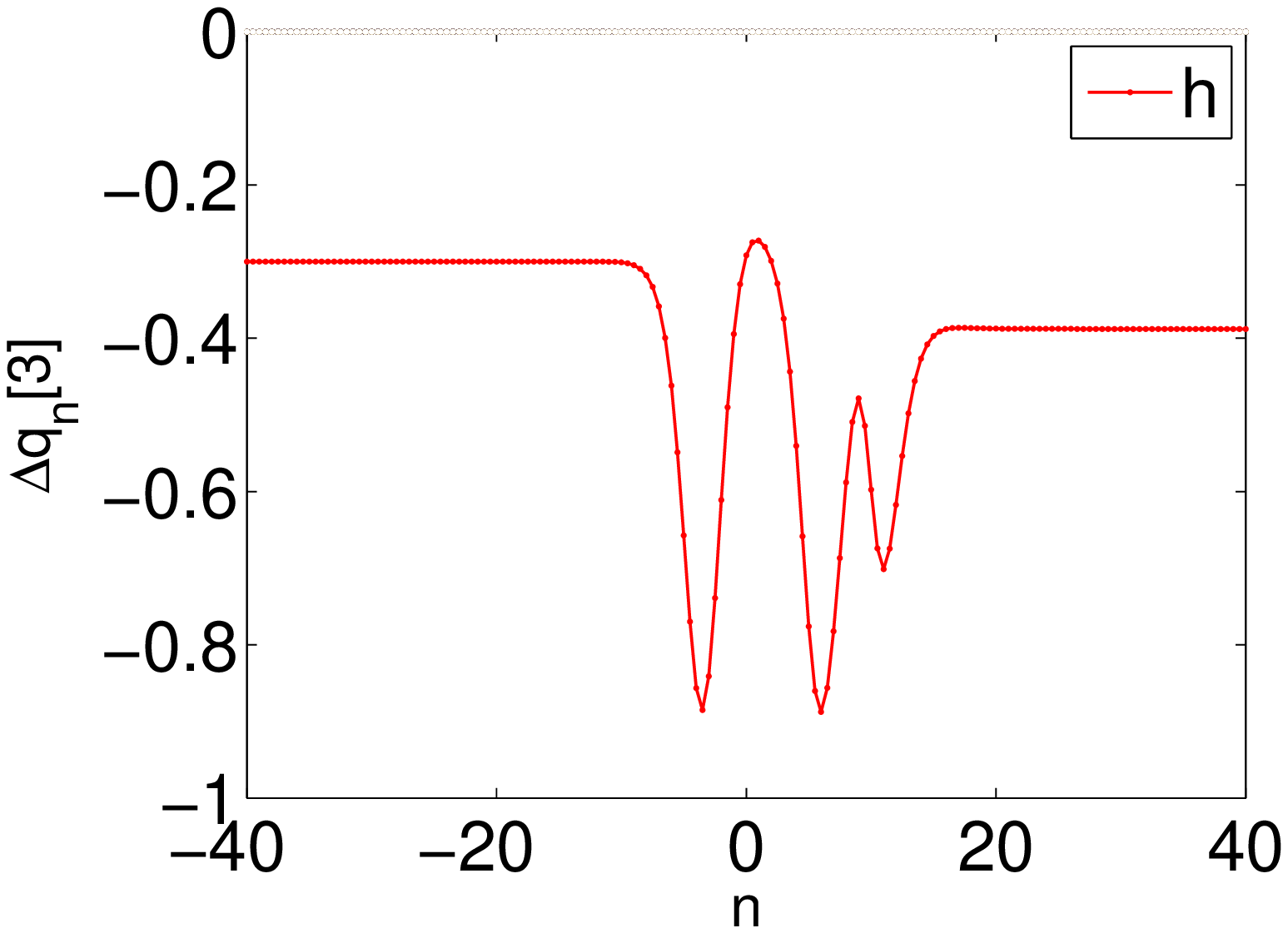}
                \label{fig:fig2}
        \end{subfigure}
        \begin{subfigure}{0.32\textwidth}
                \includegraphics[width=\textwidth]{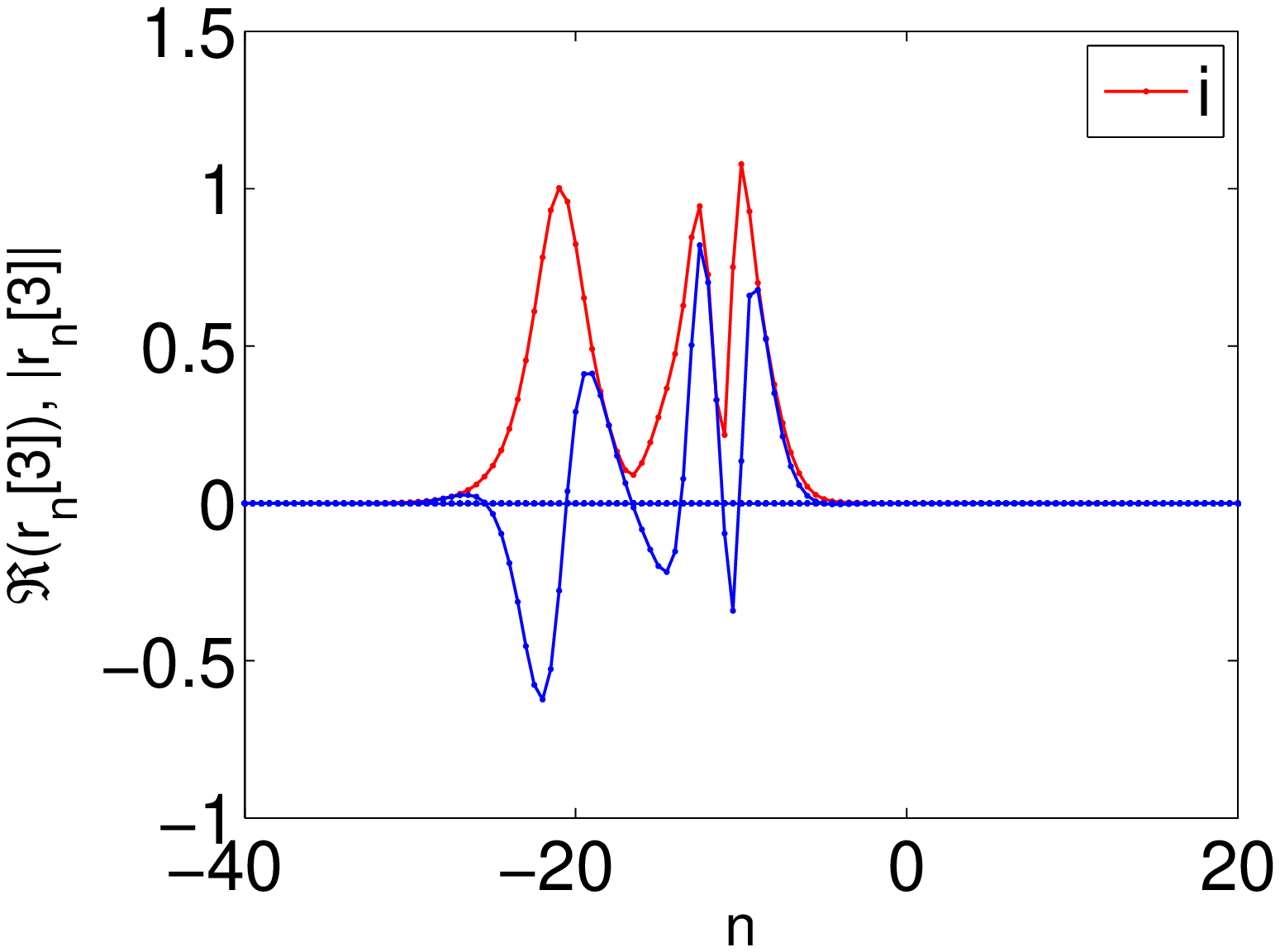}
                \label{fig:fig1}
        \end{subfigure}
        \caption{Three-kink solution in figure (g) with
        $t=20 ,\;\varrho=-1,\;\lambda_{1}=3 + \dot\imath,\;\lambda_{2}=1 + \dot\imath,\;\lambda_{3}=-0.1-\dot\imath$, dark soliton in (h) with $t=2 ,\;\varrho=-0.3,\;\lambda_{1}=0.3 + \dot\imath,\;\lambda_{2}=0.1 - \dot\imath,\;\lambda_{3}=-1-\dot\imath$
        and in figure (i); red line: $|r_{n}[3]|$, blue line $\Re{(r_{n}[3])}$  with
        $t=12,\;\varrho=0.6,\;\lambda_{1}=0.8-\dot\imath,\; \lambda_{2}=0.5+\dot\imath,\;\lambda_{3}=-0.5-\dot\imath$.}
        \label{fig:figure3}
\end{figure}

For a complex sdCD system, we have $\mathcal Q_{n} = q_{n}$ and
$\mathcal R_{n} = r_{n}$ the matrix $\mathcal U_{n}$ takes the form
\begin{equation}\label{u2by2}
\mathcal U_{n} = -\dot{\imath} \left(
\begin{array}{cc}
q_{n} & r_{n} \\
\bar{r}_{n} & -q_{n}
\end{array}%
\right).
\end{equation}
The matrix $\mathcal F_{n}$ for the complex sdCD system has the form
\begin{equation}
\mathcal F_{n} = \left(\left|f^{(1)}\right\rangle_{n},\;
\left|f^{(2)}\right\rangle_{n}\right) = \left(
\begin{array}{cc}
f^{(1)}_{n,\;11} & f^{(2)}_{n,\;12} \\
f^{(1)}_{n,\;21} & f^{(2)}_{n,\;22}
\end{array}%
\right),
\end{equation}
so that particular matrix solutions $\mathcal F_{n,\;k}$ to the Lax
pair (\ref{scalarlax1})-(\ref{scalarlax2}) at the eigenvalue
matrices $\Lambda_{k}$  are written as
\begin{equation}
\mathcal F_{n,\;k} = \left(
\begin{array}{cc}
f^{(2k-1)}_{n,\;11} & f^{(2k)}_{n,\;12} \\
f^{(2k-1)}_{n,\;21} & f^{(2k)}_{n,\;22}
\end{array}%
\right), \quad \Lambda_{k} = \left(
\begin{array}{cc}
\lambda_{2k-1} & 0 \\
0 & \lambda_{2k}
\end{array}%
\right),\quad k=1,\;2,\;...,\;K.
\end{equation}
And the $2 \times 2$ matrix $\Theta_{n}^{(K)}$ is
\begin{equation}\label{theta2by2}
\Theta_{n}^{(K)} = \left(
\begin{array}{cc}
\Theta^{(K)}_{n,\;11} & \Theta^{(K)}_{n,\;12} \\
\Theta^{(K)}_{n,\;21} & \Theta^{(K)}_{n,\;22}
\end{array}%
\right) = \left\vert
\begin{array}{cc}
\mathcal F_{n} & \mathcal E^{(K)} \\
\mathcal {\widehat F}_{n} & \fbox{$ O $}
\end{array}%
\right \vert.
\end{equation}
In the present case, $\mathcal E^{(K)}$ are $2K \times 2$ and
$\mathcal {\widehat F}_{n},\; \mathcal F_{n}$ are the $ 2 \times
2K,\; 2K \times 2K$ matrices respectively. From the equations
(\ref{SS1}) and (\ref{u2by2}) with (\ref{theta2by2}), the $K$-fold
DT on the scalar fields $q_{n}$ and $r_{n}$ are given by
\begin{eqnarray}
q_{n}^{[K]} &=& q_{n} + \dot\imath \Theta^{(K)}_{n,\;11},
\label{scalar q1} \\
r_{n}^{[K]} &=&  \dot\imath \Theta^{(K)}_{n,\;12}. \label{scalar r1}
\end{eqnarray}
For one soliton $K=1$, we have
\begin{equation}
\mathcal E^{(1)} = \left(
\begin{array}{cc}
1 & 0 \\
0 & 1
\end{array}%
\right), \quad \mathcal F_{n,\;1} = \left(
\begin{array}{cc}
f^{(1)}_{n,\;11} & f^{(2)}_{n,\;12} \\
f^{(1)}_{n,\;21} & f^{(2)}_{n,\;22}
\end{array}%
\right), \quad \Lambda_{1} = \left(
\begin{array}{cc}
\lambda_{1} & 0 \\
0 & \lambda_{2}
\end{array}%
\right),
\end{equation}
so that the matrices $\mathcal F_{n}$ and $ \widehat{\mathcal
F}_{n}$
\begin{equation}
\mathcal F_{n} = \mathcal F_{n,\;1} = \left(
\begin{array}{cc}
f^{(1)}_{n,\;11} & f^{(2)}_{n,\;12} \\
f^{(1)}_{n,\;21} & f^{(2)}_{n,\;22}
\end{array}%
\right), \quad \widehat{\mathcal F}_{n} = \mathcal F_{n,\;1}
\Lambda_{1} =  \left(
\begin{array}{cc}
\lambda_{1} f^{(1)}_{n,\;11} & \lambda_{2} f^{(2)}_{n,\;12} \\
\lambda_{1} f^{(1)}_{n,\;21} & \lambda_{2} f^{(2)}_{n,\;22}
\end{array}%
\right).
\end{equation}
Therefore, the matrix element $\Theta^{(1)}_{n,\;11}$ in the matrix
$\Theta^{(1)}_{n}$ can be computed as
\begin{eqnarray}
\Theta_{n,\;11}^{(1)} &=& \left\vert
\begin{array}{cc}
\mathcal F_{n} & \mathcal E^{(1)}_{1} \\
\mathcal ({\widehat F}_{n})_{1} & \fbox{$ O $}
\end{array}%
\right \vert = \left\vert
\begin{array}{ccc}
 f^{(1)}_{n,\;11} &  f^{(2)}_{n,\;12} & 1 \\
 f^{(1)}_{n,\;21} &  f^{(2)}_{n,\;22} & 0
\\
\lambda_{1} f^{(1)}_{n,\;11} & \lambda_{2} f^{(2)}_{n,\;12} &
\fbox{$ 0 $}
\end{array}%
\right \vert,  \\ &=& - \frac{\det \left(
\begin{array}{cc}
\lambda_{1} f^{(1)}_{n,\;11} & \lambda_{2} f^{(2)}_{n,\;12} \\
 f^{(1)}_{n,\;21} &  f^{(2)}_{n,\;22}
\end{array}%
\right)}{\det \left(
\begin{array}{cc}
 f^{(1)}_{n,\;11} &  f^{(2)}_{n,\;12} \\
 f^{(1)}_{n,\;21} &  f^{(2)}_{n,\;22}
\end{array}%
\right)}  = -\frac{\lambda_{1} f^{(1)}_{n,\;11}
f^{(2)}_{n,\;22}-\lambda_{2}f^{(1)}_{n,\;21}f^{(2)}_{n,\;12}}{
f^{(1)}_{n,\;11} f^{(2)}_{n,\;22} -
f^{(1)}_{n,\;21}f^{(2)}_{n,\;12}}. \notag
\end{eqnarray}
Let us take $f^{(2)}_{n,\;22} = \bar{f}^{(1)}_{n,\;11},\;
f^{(2)}_{n,\;12} = -\bar{f}^{(1)}_{n,\;21}$ and $\lambda_{2} =
\bar{\lambda}_{1}$, we obtain
\begin{equation}\label{the11}
\Theta_{n,\;11}^{(1)}  = -\frac{\lambda_{1}
|f^{(1)}_{n,\;11}|^{2}+\bar{\lambda}_{1}|f^{(1)}_{n,\;21}|^{2}}{
|f^{(1)}_{n,\;11}|^{2} + |f^{(1)}_{n,\;21}|^{2}}.
\end{equation}
Similarly
\begin{equation}\label{the12}
\Theta_{n,\;12}^{(1)}  = \frac{(\bar{\lambda}_{1}-\lambda_{1})
f^{(1)}_{n,\;11}|\bar{f}^{(1)}_{n,\;21}}{ |f^{(1)}_{n,\;11}|^{2} +
|f^{(1)}_{n,\;21}|^{2}}.
\end{equation}
For $\Lambda_{1} = \text{diag}(\lambda_{1},\;\bar{\lambda}_{1})$, we
have the particular matrix solution $\mathcal F_{n,\;1}$ to the Lax
pair (\ref{scalarlax1})-(\ref{scalarlax2}) as
\begin{equation}\label{f1}
\mathcal F_{n,\;1} = \left(
\begin{array}{cc}
 f^{(1)}_{n,\;11} &  -\bar{f}^{(1)}_{n,\;21} \\
 f^{(1)}_{n,\;21} &  \bar{f}^{(1)}_{n,\;11}
\end{array}%
\right) = \left(
\begin{array}{cc}
 \left(1-\dot{\imath}\varrho
\lambda^{-1}_{1}\right)^{n}e^{\frac{\dot\imath \lambda_{1}}{2}t} &
-\left(1-\dot{\imath}\varrho
\bar{\lambda}^{-1}_{1}\right)^{n}e^{\frac{\dot\imath \bar{\lambda}_{1}}{2}t} \\
 \left(1+\dot{\imath}\varrho
\lambda^{-1}_{1}\right)^{n}e^{-\frac{\dot\imath \lambda_{1}}{2}t} &
\left(1+\dot{\imath}\varrho
\bar{\lambda}^{-1}_{1}\right)^{n}e^{-\frac{\dot\imath
\bar{\lambda}_{1}}{2}t}
\end{array}%
\right).
\end{equation}
Now substituting equation (\ref{f1}) into equations (\ref{scalar
q1})-(\ref{scalar r1}) with (\ref{the11})-(\ref{the12}) yields the
one-soliton solution of the complex sdCD system, given by
\begin{eqnarray}
q_{n}^{[1]} &=& q_{n} - \dot\imath \frac{\lambda_{1} \chi_{n}^{+} +
\bar{\lambda}_{1}\chi_{n}^{-}}{ \chi_{n}^{+} + \chi_{n}^{-}},
\label{sonesolitonq1} \\
r_{n}^{[1]} &=&  \dot\imath \frac{
(\bar{\lambda}_{1}-\lambda_{1})\varphi_{n}^{+}}{ \chi_{n}^{+} +
\chi_{n}^{-}}, \label{sonesolitonr1}
\end{eqnarray}
where
\begin{eqnarray}
\chi_{n}^{+} &=& \left(1-\dot{\imath}\varrho
\lambda^{-1}_{1}\right)^{n} \left(1+\dot{\imath}\varrho
\bar{\lambda}^{-1}_{1}\right)^{n} e^{\frac{\dot\imath}{2}(
\lambda_{1}- \bar{\lambda}_{1})t}, \quad \chi_{n}^{-} =
\left(1+\dot{\imath}\varrho \lambda^{-1}_{1}\right)^{n}
\left(1-\dot{\imath}\varrho \bar{\lambda}^{-1}_{1}\right)^{n}
e^{-\frac{\dot\imath}{2}( \lambda_{1}- \bar{\lambda}_{1})t}, \notag
\\
\varphi_{n}^{+} &=& \left(1-\dot{\imath}\varrho
\lambda^{-1}_{1}\right)^{n} \left(1-\dot{\imath}\varrho
\bar{\lambda}^{-1}_{1}\right)^{n} e^{\frac{\dot\imath}{2}(
\lambda_{1}+ \bar{\lambda}_{1})t}. \label{functions}
\end{eqnarray}
The plot of the solutions is depicted in figure \ref{fig:figure1}
For two soliton, we take the matrices $\mathcal E^{(2)},\;\mathcal
F_{n,\;1},\; \mathcal F_{n,\;2},\;\Lambda_{1}$ and $\Lambda_{2}$ to
be
\begin{eqnarray}\label{c5}
&& \mathcal{E}^{(2)}=\left(\begin{array}{cc}
0 & 0 \\
0 & 0 \\
1 & 0 \\
0 & 1 \\
\end{array}%
\right), \notag \\
&& \mathcal{F}_{n,\;1}=\left(
\begin{array}{cc}
 f^{(1)}_{n,\;11} &  f^{(2)}_{n,\;12} \\
 f^{(1)}_{n,\;21} &  f^{(2)}_{n,\;22}
\end{array}%
\right), \quad \mathcal{F}_{n,\;2}=\left(
\begin{array}{cc}
 f^{(3)}_{n,\;11} &  f^{(4)}_{n,\;12} \\
 f^{(3)}_{n,\;21} &  f^{(4)}_{n,\;22}
\end{array}%
\right), \notag \\
&&\Lambda_{1}=\left(\begin{array}{cc}
\lambda_{1} & 0 \\
0 & {\lambda}_{2} \\
\end{array}%
\right), \quad \qquad \Lambda_{2}=\left(\begin{array}{cc}
                  \lambda_{3} & 0 \\
                  0 & {\lambda}_{4} \\
                \end{array}%
              \right),
\end{eqnarray}
so that the matrices $\mathcal F_{n}$ and $\widehat{\mathcal F}_{n}$
become
\begin{eqnarray}\label{c6}
\mathcal{F}_{n}&=&\left( \begin{array}{cc} \mathcal{F}_{n,\;1} &  \mathcal{F}_{n,\;2} \\
 \mathcal{F}_{n,\;1}\Lambda_{1} & \mathcal{F}_{n,\;2}\Lambda_{2}
\end{array} \right)
=\left( \begin{array}{c:c}
\begin{array}{cc}
 f^{(1)}_{n,\;11} &  f^{(2)}_{n,\;12} \\
 f^{(1)}_{n,\;21} &  f^{(2)}_{n,\;22}
\end{array} & \begin{array}{cc}
  f^{(3)}_{n,\;11} &  f^{(4)}_{n,\;12} \\
 f^{(3)}_{n,\;21} &  f^{(4)}_{n,\;22}
              \end{array} \\  \hdashline
              \begin{array}{cc}
  \lambda_{1}f^{(1)}_{n,\;11} & \lambda_{2}f^{(2)}_{n,\;12} \\
\lambda_{1}f^{(1)}_{n,\;21} & \lambda_{2}f^{(2)}_{n,\;22} \\
\end{array} &
\begin{array}{cc}
  \lambda_{3}f^{(3)}_{n,\;11} & \lambda_{4}f^{(4)}_{n,\;12} \\
\lambda_{3}f^{(3)}_{n,\;21} & \lambda_{4}f^{(4)}_{n,\;22} \\
\end{array}
\end{array}%
\right), \notag \\
\widehat{\mathcal{F}}_{n}&=&\left( \begin{array}{cc} \mathcal
F_{n,\;1}\Lambda_{1}^{2} & \mathcal F_{n,\;2}\Lambda_{2}^{2}
\end{array} \right) =\left( \begin{array}{c:c}
\begin{array}{cc}
  \lambda_{1}^{2}f^{(1)}_{n,\;11} & \lambda_{2}^{2} f^{(2)}_{n,\;12} \\
\lambda_{1}^{2}f^{(1)}_{n,\;21} & \lambda_{2}^{2} f^{(2)}_{n,\;22} \\
\end{array} &
\begin{array}{cc}
  \lambda_{3}^{2}f^{(3)}_{n,\;11} & \lambda_{4}^{2}f^{(4)}_{n,\;12} \\
\lambda_{3}^{2}f^{(3)}_{n,\;21} & \lambda_{4}^{2}f^{(4)}_{n,\;22} \\
\end{array}
\end{array}%
\right). \notag
\end{eqnarray}
The two-fold DT on the scalar fields $q_{n}$ and $r_{n}$ is given by
\begin{eqnarray}
q_{n}^{[2]} &=& q_{n} + \dot\imath \Theta^{(2)}_{n,\;11},
\label{scalar q2} \\
r_{n}^{[2]} &=&  \dot\imath \Theta^{(2)}_{n,\;12}, \label{scalar r2}
\end{eqnarray}
where the matrix elements
$\Theta^{(2)}_{n,\;11},\;\Theta^{(2)}_{n,\;12}$ can be computed as
ratios of determinants
\begin{eqnarray}\label{the11 twofold}
\Theta_{n,\;11}^{(2)} &=& \left\vert
\begin{array}{cc}
\mathcal F_{n} & \mathcal E^{(2)}_{1} \\
(\widehat{\mathcal F}_{n})_{1} & \fbox{$ O $}
\end{array}%
\right \vert = \left\vert  \begin{array}{ccccc}
 f^{(1)}_{n,\;11} &  f^{(2)}_{n,\;12} & f^{(3)}_{n,\;11} &  f^{(4)}_{n,\;12} & 0 \\
 f^{(1)}_{n,\;21} &  f^{(2)}_{n,\;22} & f^{(3)}_{n,\;21} &
 f^{(4)}_{n,\;22} & 0 \\
  \lambda_{1}f^{(1)}_{n,\;11} & \lambda_{2}f^{(2)}_{n,\;12} & \lambda_{3}f^{(3)}_{n,\;11} & \lambda_{4}f^{(4)}_{n,\;12} & 1 \\
\lambda_{1}f^{(1)}_{n,\;21} & \lambda_{2}f^{(2)}_{n,\;22} & \lambda_{3}f^{(3)}_{n,\;21} & \lambda_{4}f^{(4)}_{n,\;22} & 0 \\
\lambda_{1}^{2}f^{(1)}_{n,\;11} & \lambda_{2}^{2} f^{(2)}_{n,\;12} &
\lambda_{3}^{2}f^{(3)}_{n,\;11} & \lambda_{4}^{2} f^{(4)}_{n,\;12} &
\fbox{$0$}
\end{array}%
 \right \vert, \notag \\ &=& - \frac{\det
\left(
\begin{array}{cccc}
 f^{(1)}_{n,\;11} &  f^{(2)}_{n,\;12} & f^{(3)}_{n,\;11} &  f^{(4)}_{n,\;12}  \\
 f^{(1)}_{n,\;21} &  f^{(2)}_{n,\;22} & f^{(3)}_{n,\;21} &
 f^{(4)}_{n,\;22}  \\
  \lambda_{1}^{2}f^{(1)}_{n,\;11} & \lambda_{2}^{2} f^{(2)}_{n,\;12} &
\lambda_{3}^{2}f^{(3)}_{n,\;11} & \lambda_{4}^{2} f^{(4)}_{n,\;12} \\
\lambda_{1}f^{(1)}_{n,\;21} & \lambda_{2}f^{(2)}_{n,\;22} &
\lambda_{3}f^{(3)}_{n,\;21} & \lambda_{4}f^{(4)}_{n,\;22}
\end{array}%
\right)}{\det \left(
\begin{array}{cccc}
 f^{(1)}_{n,\;11} &  f^{(2)}_{n,\;12} & f^{(3)}_{n,\;11} &  f^{(4)}_{n,\;12}  \\
 f^{(1)}_{n,\;21} &  f^{(2)}_{n,\;22} & f^{(3)}_{n,\;21} &
 f^{(4)}_{n,\;22}  \\
  \lambda_{1}f^{(1)}_{n,\;11} & \lambda_{2}f^{(2)}_{n,\;12} & \lambda_{3}f^{(3)}_{n,\;11} & \lambda_{4}f^{(4)}_{n,\;12} \\
\lambda_{1}f^{(1)}_{n,\;21} & \lambda_{2}f^{(2)}_{n,\;22} &
\lambda_{3}f^{(3)}_{n,\;21} & \lambda_{4}f^{(4)}_{n,\;22}
\end{array}%
\right)}.
\end{eqnarray}
Similarly, the matrix elements $\Theta_{n,\;12}^{(2)}$ are
\begin{eqnarray}\label{the12 twofold}
\Theta_{n,\;12}^{(2)} &=& - \frac{\det \left(
\begin{array}{cccc}
 f^{(1)}_{n,\;11} &  f^{(2)}_{n,\;12} & f^{(3)}_{n,\;11} &  f^{(4)}_{n,\;12}  \\
 f^{(1)}_{n,\;21} &  f^{(2)}_{n,\;22} & f^{(3)}_{n,\;21} &
 f^{(4)}_{n,\;22}  \\
  \lambda_{1}f^{(1)}_{n,\;11} & \lambda_{2}f^{(2)}_{n,\;12} & \lambda_{3}f^{(3)}_{n,\;11} & \lambda_{4}f^{(4)}_{n,\;12} \\
\lambda_{1}^{2}f^{(1)}_{n,\;11} & \lambda_{2}^{2} f^{(2)}_{n,\;12} &
\lambda_{3}^{2}f^{(3)}_{n,\;11} & \lambda_{4}^{2} f^{(4)}_{n,\;12}
\end{array}%
\right)}{\det \left(
\begin{array}{cccc}
 f^{(1)}_{n,\;11} &  f^{(2)}_{n,\;12} & f^{(3)}_{n,\;11} &  f^{(4)}_{n,\;12}  \\
 f^{(1)}_{n,\;21} &  f^{(2)}_{n,\;22} & f^{(3)}_{n,\;21} &
 f^{(4)}_{n,\;22}  \\
  \lambda_{1}f^{(1)}_{n,\;11} & \lambda_{2}f^{(2)}_{n,\;12} & \lambda_{3}f^{(3)}_{n,\;11} & \lambda_{4}f^{(4)}_{n,\;12} \\
\lambda_{1}f^{(1)}_{n,\;21} & \lambda_{2}f^{(2)}_{n,\;22} &
\lambda_{3}f^{(3)}_{n,\;21} & \lambda_{4}f^{(4)}_{n,\;22}
\end{array}%
\right)}.
\end{eqnarray}
By substituting equations (\ref{the11 twofold})-(\ref{the12
twofold}) into equations (\ref{scalar q2})-(\ref{scalar r2})
respectively, we get the two-fold DT on the fields $q_{n}$ and
$r_{n}$. Further, we use $f^{(2l)}_{n,\;22} =
(-1)^{2l}\bar{f}^{(2l-1)}_{n,\;11},\; f^{(2l)}_{n,\;12} =
(-1)^{2l-1}\bar{f}^{(2l-1)}_{n,\;21}$ and $\lambda_{2l} =
\bar{\lambda}_{2l-1}$ where $l=1,\;2$ in equations (\ref{scalar q2})
and (\ref{scalar r2}) to obtain two-soliton solutions of the complex
sdCD system. These solutions are plotted in figure \ref{fig:figure2}
To get three-soliton solution, we take three particular matrix
solutions $\mathcal F_{n,\;k}$ with the eigenvalue matrices
$\Lambda_{k}$ ($k=1,\;2,\;3$). With these particular solutions, we
obtain three soliton solutions depicted in figure \ref{fig:figure3}
Now, we would like to reduce the semi-discrete solutions of complex
sdCD system to those of continuous solutions of the complex CD
system by applying continuum limit. For this, replace $\varrho
\rightarrow \delta \varrho$ and send $\delta$ to zero,  then
equations (\ref{sonesolitonq1})-(\ref{sonesolitonr1}) respectively
reduce to the form
\begin{eqnarray}
q^{[1]} &=& q - \dot\imath \left(\lambda_{1R} + \dot\imath
\lambda_{1I} \tanh a\right), \label{contonesolitonq1} \\
r^{[1]} &=& \lambda_{1I} e^{\dot\imath b} \textrm{sech} a,
\label{contonesolitonr1}
\end{eqnarray}
where
\begin{equation}
a= -\lambda_{1I} \left(\frac{2
\varrho}{\left|\lambda_{1}\right|^{2}}x + t\right), \qquad b=
-\lambda_{1R} \left(\frac{2 \varrho}{\left|\lambda_{1}\right|^{2}}x
- t\right).
\end{equation}
Equations (\ref{contonesolitonq1})-(\ref{contonesolitonr1})
represents one-soliton solution of the complex CD system
(\ref{usualCD1})-(\ref{usualCD2}). Similarly, one can also find the
two- and three soliton solutions of the complex CD system
(\ref{usualCD1})-(\ref{usualCD2}) by applying continuum the limit on
the solutions as obtained for the complex sdCD system.
\subsection{2-component complex sdCD system}
For 2-component complex sdCD system, the $4 \times 4$ matrix
$\mathcal U_{n}$ takes the form
\begin{equation}\label{umatrixtwocomp}
\mathcal U_{n} = -\dot\imath\left(\begin{array}{cccc} q_{n} & 0 &
r_{n}^{(1)} & r_{n}^{(2)} \\
0 & q_{n} & -\bar{r}_{n}^{(2)} & \bar{r}_{n}^{(1)} \\
\bar{r}_{n}^{(1)} & -r_{n}^{(2)} & -q_{n} & 0 \\
\bar{r}_{n}^{(2)} & r_{n}^{(1)} & 0 & -q_{n}
\end{array}%
\right).
\end{equation}
The matrix $\Theta_{n}^{(K)}$ in equation (\ref{S2}) becomes
\begin{equation}\label{theta4by4}
\Theta_{n}^{(K)} = \left( \begin{array}{c:c}
\begin{array}{cc}
 \Theta^{(K)}_{n,\;11} &  \Theta^{(K)}_{n,\;12} \\
 \Theta^{(K)}_{n,\;21} &  \Theta^{(K)}_{n,\;22}
\end{array} & \begin{array}{cc}
  \Theta^{(K)}_{n,\;13} &  \Theta^{(K)}_{n,\;14} \\
 \Theta^{(K)}_{n,\;23} &  f^{(K)}_{n,\;24}
              \end{array} \\  \hdashline
              \begin{array}{cc}
  \Theta^{(K)}_{n,\;31} &  \Theta^{(K)}_{n,\;32} \\
 \Theta^{(K)}_{n,\;41} &  \Theta^{(K)}_{n,\;42}
\end{array} & \begin{array}{cc}
  \Theta^{(K)}_{n,\;33} &  \Theta^{(K)}_{n,\;34} \\
 \Theta^{(K)}_{n,\;43} &  f^{(K)}_{n,\;44}
              \end{array}
\end{array}%
\right) = \left\vert
\begin{array}{cc}
\mathcal F_{n} & \mathcal E^{(K)} \\
\mathcal {\widehat F}_{n} & \fbox{$ O $}
\end{array}%
\right \vert.
\end{equation}
In this case, $\mathcal E^{(K)}$ are $4K \times 4$ and $\mathcal
{\widehat F}_{n},\; \mathcal F_{n}$ are the $ 4 \times 4K,\; 4K
\times 4K$ matrices respectively. From equations (\ref{SS1}) and
(\ref{umatrixtwocomp})-(\ref{theta4by4}), the $K$-fold DT on the
solutions of the 2-component complex sdCD system is given by
\begin{eqnarray}
q_{n}^{[K]} &=& q_{n} + \dot\imath \Theta_{n,\;11}^{(K)},
\label{q1twocomp} \\
r_{n}^{(1)[K]} &=&  \dot\imath \Theta_{n,\;13}^{(K)},
\label{r1twocomp} \\
r_{n}^{(2)[K]} &=&  \dot\imath \Theta_{n,\;14}^{(K)},
\label{r2twocomp}
\end{eqnarray}

\begin{figure}[H]
        \centering
        \begin{subfigure}[l]{0.35\textwidth}
                \includegraphics[width=\textwidth]{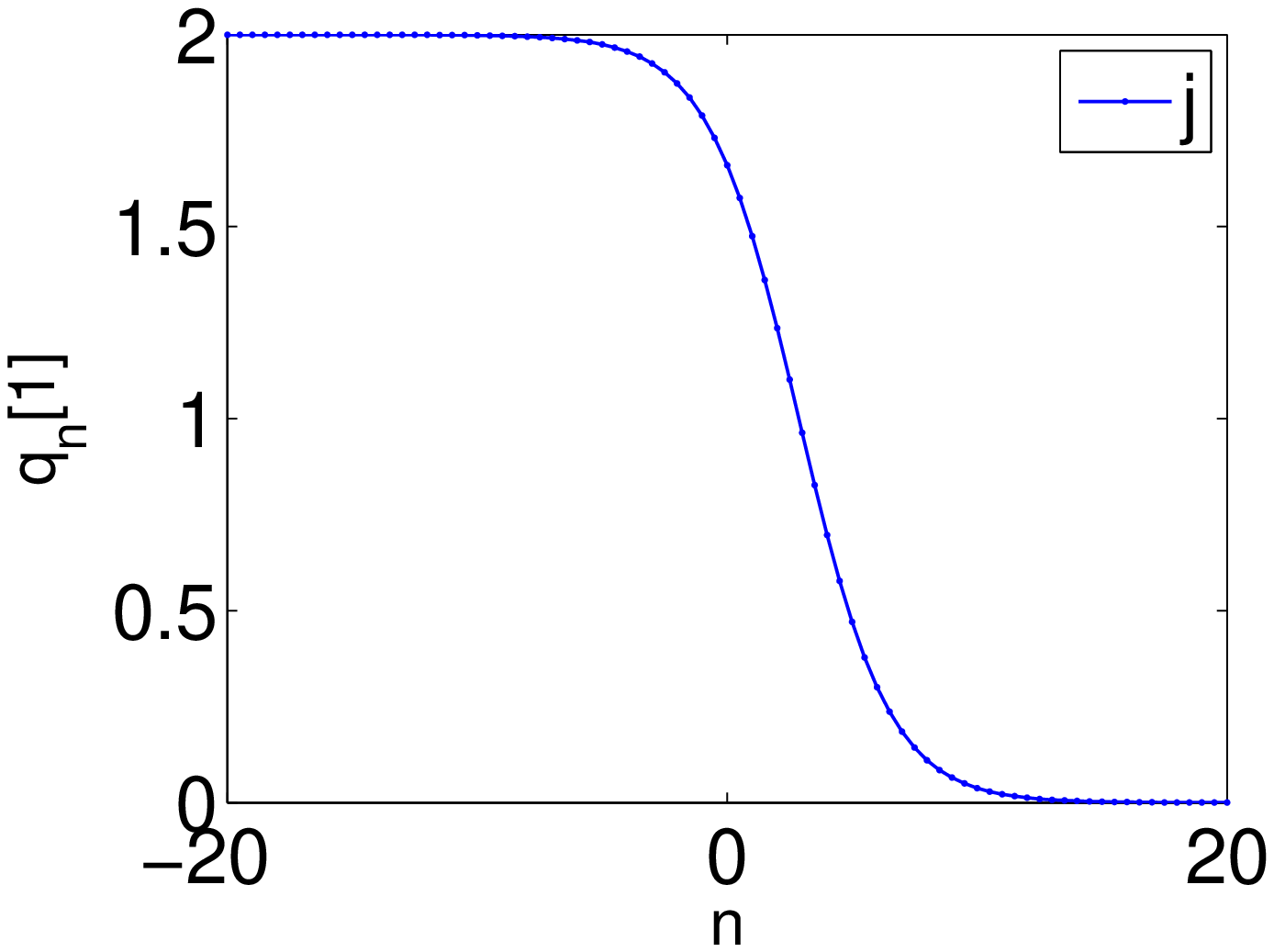}
                \label{fig:fig1}
        \end{subfigure}
         \begin{subfigure}[r]{0.35\textwidth}
                \includegraphics[width=\textwidth]{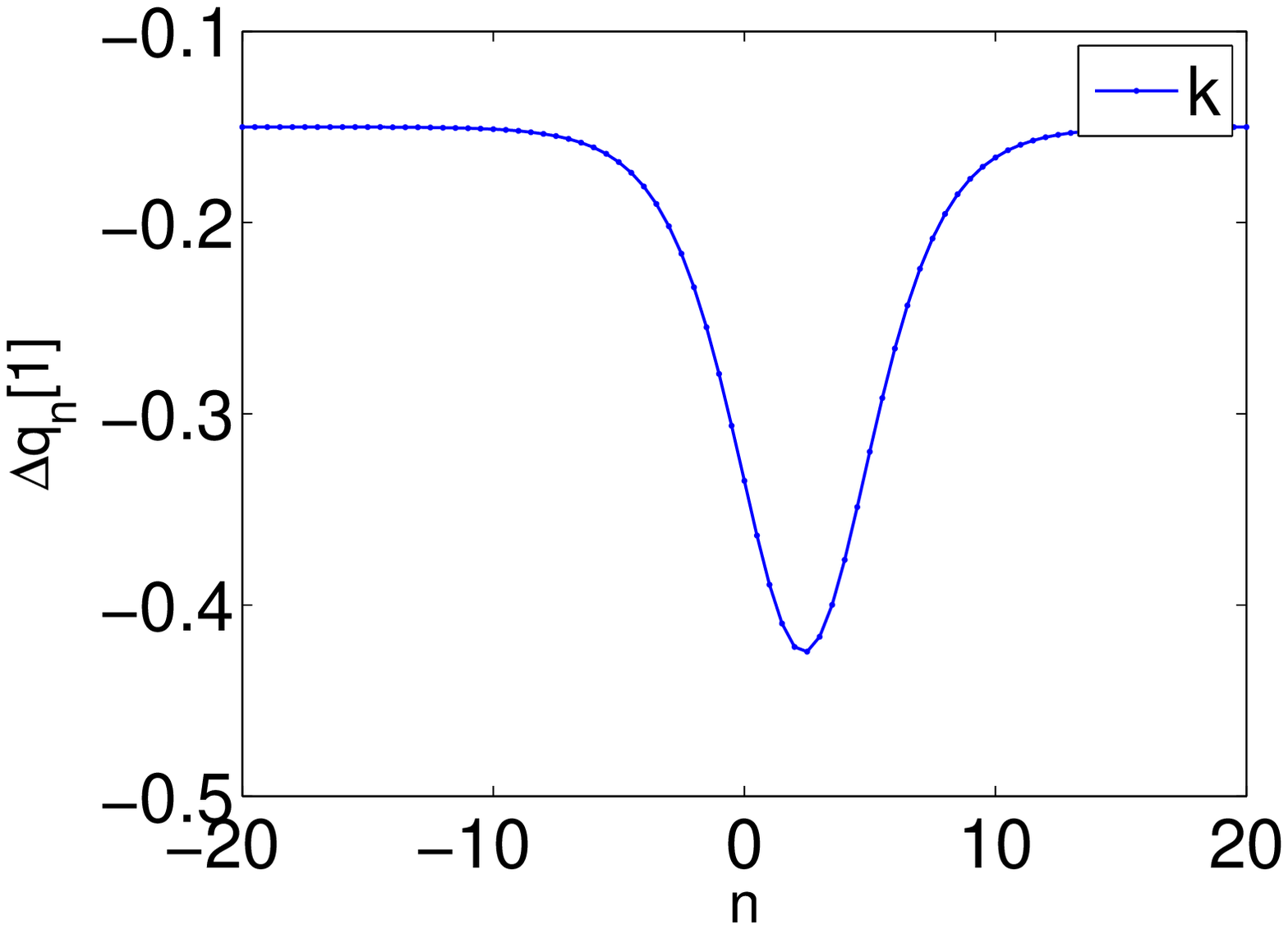}
                \label{fig:fig2}
        \end{subfigure}
        \caption{Kink and dark solutions in figures (j and k) with $t=0.1,\;\varrho=0.15,\;\lambda_{1}=0.3-\dot\imath,\;\epsilon_{1}=1-\dot\imath,\;\epsilon_{2}=1+\dot\imath$.}
        \label{fig:figure8}
\end{figure}

\begin{figure}[H]
        \centering
        \begin{subfigure}[l]{0.35\textwidth}
                \includegraphics[width=\textwidth]{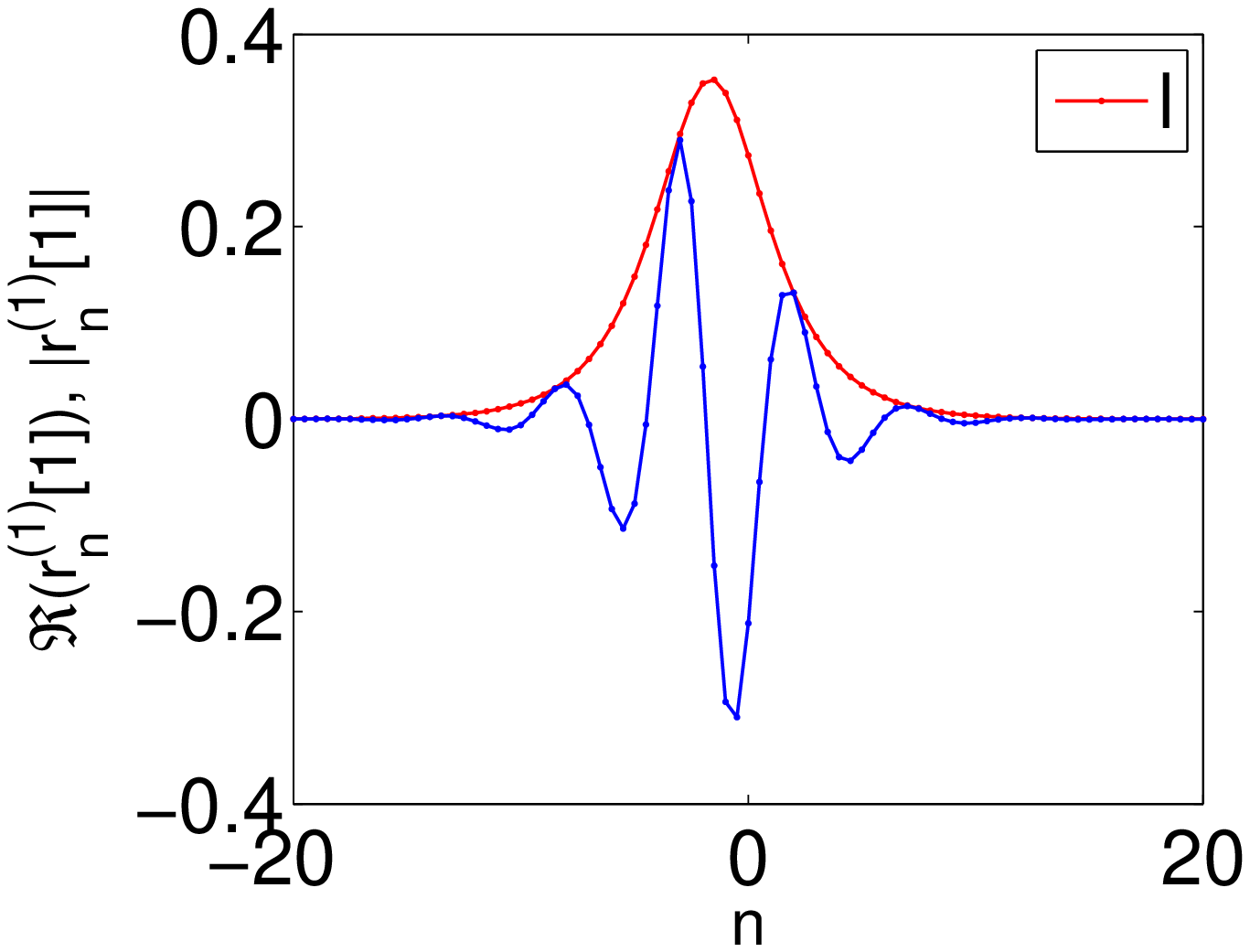}
                \label{fig:fig1}
        \end{subfigure}
        \begin{subfigure}[r]{0.35\textwidth}
                \includegraphics[width=\textwidth]{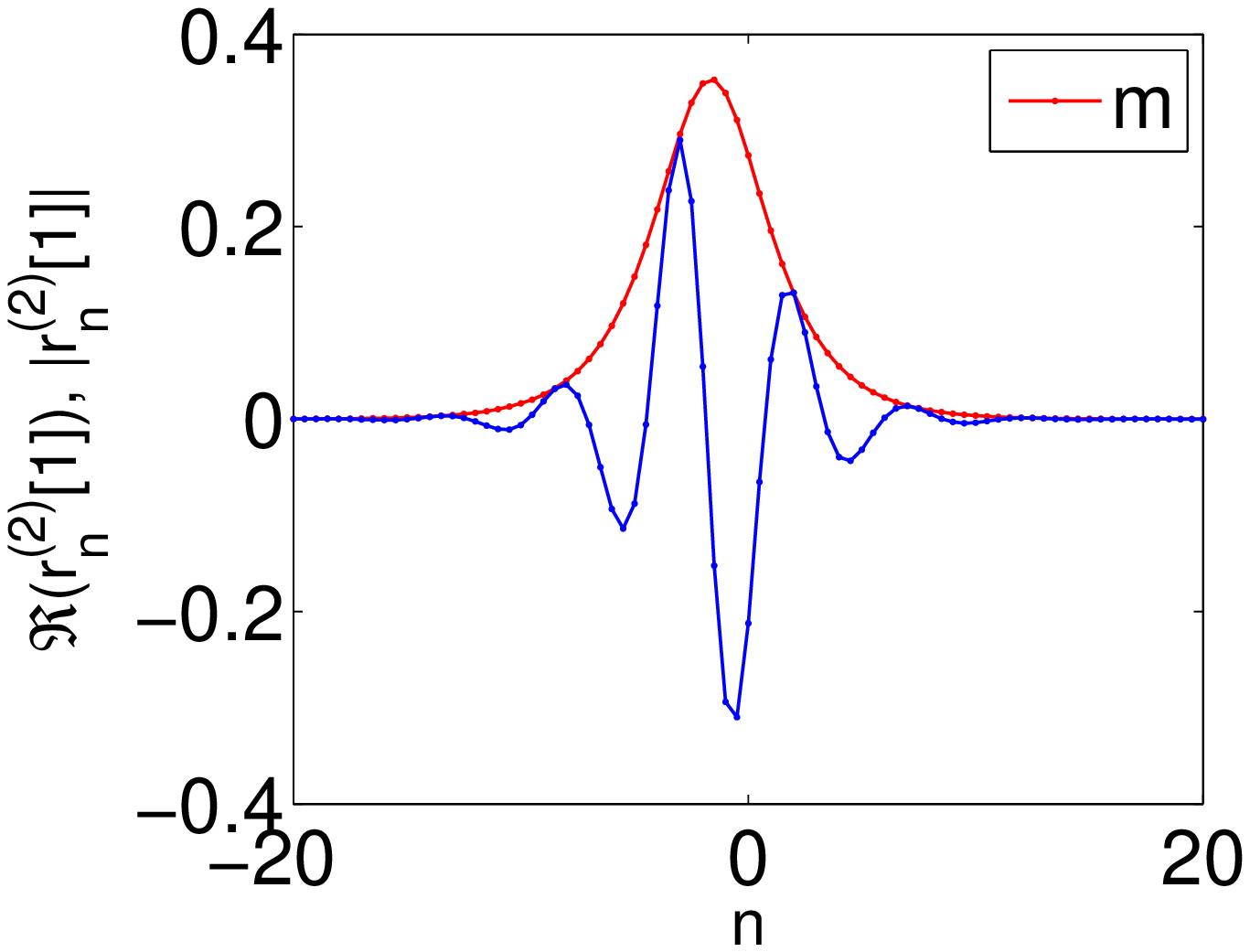}
                \label{fig:fig1}
        \end{subfigure}
        \caption{Bright and periodic solutions in figures (l and m); red line: $|r_{n}|$, blue line $\Re{(r_{n})}$ with $t=0.1,\;\varrho=0.8,\;\lambda_{1}=1-0.5\dot\imath,\;\epsilon_{1}=1-\dot\imath,\;\epsilon_{2}=1+\dot\imath$.}
        \label{fig:figure9}
\end{figure}

The matrix valued solution $\Psi_{n}$ to the Lax pair of the
2-component sdCD system is written as
\begin{equation}\label{matrixseed twocomp}
\Psi_{n} = \left(
\begin{array}{c:c}
\left(1-\dot{\imath}\varrho
\lambda^{-1}\right)^{n}e^{\frac{\dot\imath \lambda}{2}t} I_{2
\times 2} & O \\
\hdashline O & \left(1+\dot{\imath}\varrho
\lambda^{-1}\right)^{n}e^{-\frac{\dot\imath \lambda}{2}t} I_{2
\times 2}
\end{array}%
\right)
\end{equation}
To construct matrix $\mathcal F_{n}$ from a matrix solution
$\Psi_{n}$, let us take columns i.e.,
\begin{eqnarray}
\left|e_{1}\right\rangle &=& \left(\begin{array}{c} \epsilon_1 \\
\epsilon_2 \\
-1 \\
0
\end{array}%
\right), \quad \left|e_{2}\right\rangle = \left(\begin{array}{c} \bar{\epsilon}_2 \\
-\bar{\epsilon}_1 \\
0 \\
1
\end{array}%
\right), \notag \\
\left|e_{3}\right\rangle &=& \left(\begin{array}{c} 1 \\
0 \\
\bar{\epsilon}_1 \\
-\epsilon_2
\end{array}%
\right), \quad \left|e_{4}\right\rangle = \left(\begin{array}{c} 0 \\
1 \\
\bar{\epsilon}_2 \\
\epsilon_1
\end{array}%
\right),
\end{eqnarray}
where $\epsilon_{1},\;\epsilon_{2}$ are complex constants. For
$\Lambda_{k} =
\text{diag}(\lambda_{k},\;\lambda_{k},\;\bar{\lambda}_{k},\;\bar{\lambda}_{k}),\;k=1,\;2,\;...,\;K$,
we have different particular matrix solutions as
\begin{eqnarray}\label{particular matrix twocomp}
\mathcal F_{n,\;k} &=&
\left(\Psi_{n}\left|e_{1}\right\rangle,\;\Psi_{n}\left|e_{2}\right\rangle,\;\Psi_{n}\left|e_{3}\right\rangle
,\;\Psi_{n}\left|e_{4}\right\rangle\right), \notag \\
&=& \left(\begin{array}{cccc} \epsilon_{1} x(\lambda_{k}) &
\bar{\epsilon}_{2} x(\lambda_{k}) & x(\bar{\lambda}_{k}) & 0 \\
\epsilon_{2} x(\lambda_{k}) & -\bar{\epsilon}_{1} x(\lambda_{k}) & 0
&  x(\bar{\lambda}_{k}) \\
- \bar{x}(\lambda_{k}) & 0 & \bar{\epsilon}_{1}
\bar{x}(\bar{\lambda}_{k}) & \bar{\epsilon}_{2}
\bar{x}(\bar{\lambda}_{k}) \\
0 & \bar{x}(\lambda_{k}) & -\epsilon_{2} \bar{x}(\bar{\lambda}_{k})
& \epsilon_{1} \bar{x}(\bar{\lambda}_{k})
\end{array}%
\right),
\end{eqnarray}
where $x(\lambda_{k}) = \left(1-\dot{\imath}\varrho
\lambda^{-1}_{k}\right)^{n}e^{\frac{\dot\imath \lambda_{k}}{2}t}$
and $\bar{x}(\lambda_{k}) = \left(1+\dot{\imath}\varrho
\lambda^{-1}_{k}\right)^{n}e^{-\frac{\dot\imath \lambda_{k}}{2}t}$.
To calculate soliton solutions explicitly of the 2-component sdCD
system, we proceed as follows. For one soliton $K=1$, the matrices
$\mathcal E^{(1)},\; \Lambda_{1},\; \mathcal F_{n},\;
\widehat{\mathcal F}_{n}$ are given by
\begin{eqnarray}\label{K1}
&&  \mathcal E^{(1)} = I_4 = \left(\begin{array}{cccc} 1 &
0 & 0 &0 \\
0 & 1 & 0 & 0\\
0& 0& 1 &0 \\
0& 0& 0& 1
\end{array}%
\right),\notag \\ && \mathcal F_{n} = \mathcal F_{n,\;1} =
\left(\begin{array}{cccc} \epsilon_{1} x(\lambda_{1}) &
\bar{\epsilon}_{2} x(\lambda_{1}) & x(\bar{\lambda}_{1}) & 0 \\
\epsilon_{2} x(\lambda_{1}) & -\bar{\epsilon}_{1} x(\lambda_{1}) & 0
&  x(\bar{\lambda}_{1}) \\
- \bar{x}(\lambda_{1}) & 0 & \bar{\epsilon}_{1}
\bar{x}(\bar{\lambda}_{1}) & \bar{\epsilon}_{2}
\bar{x}(\bar{\lambda}_{1}) \\
0 & \bar{x}(\lambda_{1}) & -\epsilon_{2} \bar{x}(\bar{\lambda}_{1})
& \epsilon_{1} \bar{x}(\bar{\lambda}_{1})
\end{array}%
\right),  \Lambda_1 = \left(\begin{array}{cccc} \lambda_{1} & 0
 & 0 & 0
 \\
 0& \lambda_1 &0 &0  \\
 0 & 0 & \bar{\lambda}_1 &0  \\
 0 & 0 & 0 & \bar{\lambda}_1
\end{array}%
\right), \notag \\
&& \widehat{\mathcal F}_{n} = \mathcal F_{n,\;1}\Lambda_1 =
\left(\begin{array}{cccc} \epsilon_{1} \lambda_{1} x(\lambda_{1}) &
\bar{\epsilon}_{2} \lambda_{1} x(\lambda_{1}) & \bar{\lambda}_{1}x(\bar{\lambda}_{1}) & 0 \\
\epsilon_{2} \lambda_{1} x(\lambda_{1}) & -\bar{\epsilon}_{1}
{\lambda}_{1} x(\lambda_{1}) & 0
& \bar{\lambda}_{1} x(\bar{\lambda}_{1}) \\
- \lambda_{1} \bar{x}(\lambda_{1}) & 0 & \bar{\epsilon}_{1}
\bar{\lambda}_{1} \bar{x}(\bar{\lambda}_{1}) & \bar{\epsilon}_{2}
\bar{\lambda}_{1} \bar{x}(\bar{\lambda}_{1}) \\
0 & \lambda_{1} \bar{x}(\lambda_{1}) &
-\epsilon_{2}\bar{\lambda}_{1} \bar{x}(\bar{\lambda}_{1}) &
\epsilon_{1} \bar{\lambda}_{1} \bar{x}(\bar{\lambda}_{1})
\end{array}%
\right).
\end{eqnarray}
By substituting (\ref{K1}) in (\ref{q1twocomp}), the one-fold DT
$q_{n}[1]$ can be calculated as
\begin{eqnarray}
{q}_{n}[1] &=& q_{n} + (\Theta_{n}^{[1]})_{11}  = q_{n} + \left\vert
\begin{array}{cc}
\mathcal F_{n,\;1} & \mathcal E^{(1)}_{1} \\
(\mathcal {\widehat F}_{n,\;1})_{1} & \fbox{$0$}%
\end{array}%
\right\vert , \notag \\
&=& q_{n} + \left\vert
\begin{array}{ccccc}
\epsilon_{1} x(\lambda_{1}) &
\bar{\epsilon}_{2} x(\lambda_{1}) & x(\bar{\lambda}_{1}) & 0 & 1\\
\epsilon_{2} x(\lambda_{1}) & -\bar{\epsilon}_{1} x(\lambda_{1}) & 0
&  x(\bar{\lambda}_{1}) & 0\\
- \bar{x}(\lambda_{1}) & 0 & \bar{\epsilon}_{1}
\bar{x}(\bar{\lambda}_{1}) & \bar{\epsilon}_{2}
\bar{x}(\bar{\lambda}_{1}) & 0\\
0 & \bar{x}(\lambda_{1}) & -\epsilon_{2} \bar{x}(\bar{\lambda}_{1})
& \epsilon_{1} \bar{1}(\bar{\lambda}_{1}) & 0 \\
\epsilon_{1} \lambda_{1} x(\lambda_{1}) & \bar{\epsilon}_{2}
\lambda_{1} x(\lambda_{1}) & \bar{\lambda}_{1}x(\bar{\lambda}_{1}) &
0& \fbox{$0$}%
\end{array}%
\right\vert, \\
&=& q_{n} - \frac{\det\left(\begin{array}{cccc} \epsilon_{1}
\lambda_{1} x(\lambda_{1}) & \bar{\epsilon}_{2} \lambda_{1}
x(\lambda_{1}) & \bar{\lambda}_{1}x(\bar{\lambda}_{1}) &
0  \\
\epsilon_{2} x(\lambda_{1}) & -\bar{\epsilon}_{1} x(\lambda_{1}) & 0
&  x(\bar{\lambda}_{1}) \\
- \bar{x}(\lambda_{1}) & 0 & \bar{\epsilon}_{1}
\bar{x}(\bar{\lambda}_{1}) & \bar{\epsilon}_{2}
\bar{x}(\bar{\lambda}_{1})  \\
0 & \bar{x}(\lambda_{1}) & -\epsilon_{2} \bar{x}(\bar{\lambda}_{1})
& \epsilon_{1} \bar{x}(\bar{\lambda}_{1})
\end{array}%
\right)}{\det\left(\begin{array}{cccc} \epsilon_{1} x(\lambda_{1}) &
\bar{\epsilon}_{2} x(\lambda_{1}) & x(\bar{\lambda}_{1}) & 0 \\
\epsilon_{2} x(\lambda_{1}) & -\bar{\epsilon}_{1} x(\lambda_{1}) & 0
&  x(\bar{\lambda}_{1}) \\
- \bar{x}(\lambda_{1}) & 0 & \bar{\epsilon}_{1}
\bar{x}(\bar{\lambda}_{1}) & \bar{\epsilon}_{2}
\bar{x}(\bar{\lambda}_{1}) \\
0 & \bar{x}(\lambda_{1}) & -\epsilon_{2} \bar{x}(\bar{\lambda}_{1})
& \epsilon_{1} \bar{1}(\bar{\lambda}_{1})
\end{array}%
\right)}. \label{DTq1}
\end{eqnarray}
Simplifying the above expression, we get
\begin{eqnarray}
{q}_{n}[1] &=& q_{n} - \dot\imath
\frac{\lambda_{1}\left(|\epsilon_{1}|^{2} +
|\epsilon_{2}|^{2}\right)\chi_{n}^{+}+\bar{\lambda}_{1}\chi_{n}^{-}}
{\left(|\epsilon_{1}|^{2} +
|\epsilon_{2}|^{2}\right)\chi_{n}^{+}+\chi_{n}^{-}}.
\label{q1soliton}
\end{eqnarray}
Similarly
\begin{eqnarray}
{r}_{n}^{(1)}[1] &=& \dot\imath\epsilon_{1}
\frac{\left(\bar{\lambda}_{1}-\lambda_{1}\right)\varphi_{n}^{+}}
{\left(|\epsilon_{1}|^{2} +
|\epsilon_{2}|^{2}\right)\chi_{n}^{+}+\chi_{n}^{-}},
\label{r1soliton} \\
{r}_{n}^{(2)}[1] &=& \dot\imath\bar{\epsilon}_{2}
\frac{\left(\bar{\lambda}_{1}-\lambda_{1}\right)\varphi_{n}^{+}}
{\left(|\epsilon_{1}|^{2} + |\epsilon_{2}|^{2}\right)\chi_{n}^{+} +
\chi_{n}^{-}}. \label{r2soliton}
\end{eqnarray}
where $\chi_{n}^{+},\; \chi_{n}^{-}$ and $\varphi_{n}^{+}$ are given
in (\ref{functions}). Equations (\ref{q1soliton})-(\ref{r2soliton})
represent one-soliton solution of the 2-component sdCD system
(\ref{twocomponent1})-(\ref{twocomponent2}). The plot of equations
(\ref{q1soliton})-(\ref{r2soliton}) has been sketched out as in the
figures \ref{fig:figure8}-\ref{fig:figure9}
\vspace{-.8cm}
And the solutions obtained in equations
(\ref{q1soliton})-(\ref{r2soliton}) in the continuum limit can be
written as
\begin{eqnarray}
q^{[1]} &=& q - \dot\imath \left(\lambda_{1R}  + \dot\imath
\tanh\left(a+ \ln\sqrt{\epsilon}\right)\right),
\label{cont2compq1soliton} \\
r^{(1)[1]} &=& \frac{\epsilon_{1}\lambda_{1I}e^{\dot\imath
b}}{\sqrt{\epsilon}}\textrm{sech}\left(a+ \ln\sqrt{\epsilon}\right),
\quad r^{(2)[1]} = \frac{\bar{\epsilon}_{2}\lambda_{1I}e^{\dot\imath
b}}{\sqrt{\epsilon}}\textrm{sech}\left(a+ \ln\sqrt{\epsilon}\right),
\label{cont2comprsoliton}
\end{eqnarray}
where $\epsilon = |\epsilon_{1}|^{2} + |\epsilon_{2}|^{2}$.
Equations (\ref{cont2compq1soliton})-(\ref{cont2comprsoliton})
represents one-soliton solutions of the 2-component complex CD
system.

\section{Concluding remarks}
In this paper, we have studied integrable discretization of complex
and multi-component coupled dispersionless system. By writing down
the Lax pair of the systems, we have computed one-, two- and
three-soliton solutions of complex and 2-component complex coupled
dispersionless system. We have also shown that, the solutions
obtained for the complex and 2-component complex sdCD system reduced
to the solutions of the respective continuous complex and
2-component complex CD system by applying continuum limit. The study
can be further extended by investigating multicomponent and matrix
generalizations of related integrable systems. An important example
of such systems is the short pulse equation. We shall address these
research problems in forthcoming work.

\end{document}